\documentclass[a4paper, 11pt]{article}
\usepackage{jheppub}
\usepackage{amssymb, amsmath}

\usepackage{graphicx}
\usepackage{wrapfig}
\usepackage[normalem]{ulem} 
\usepackage{hyperref}
\usepackage[pdftex]{pict2e}
\pdfoutput=1
\title{An Effective Field Theory for Large Oscillons}
\author[a, b]{D.G.~Levkov,}
\author[a, b, c]{V.E.~Maslov,} 
\author[a]{E.Ya.~Nugaev,}
\author[a]{and A.G.~Panin}
\affiliation[a]{\footnotesize Institute for Nuclear Research of the
  Russian Academy of Sciences, Moscow 117312, Russia}
\affiliation[b]{\footnotesize Institute for Theoretical and Mathematical
  Physics, MSU, Moscow 119991, Russia}
\affiliation[c]{\footnotesize{Department of Particle Physics and Cosmology, Faculty of Physics, MSU, Moscow 119991, Russia}}
\emailAdd{levkov@ms2.inr.ac.ru}
\emailAdd{vasilevgmaslov@ms2.inr.ac.ru}
\emailAdd{emin@ms2.inr.ac.ru}
\emailAdd{panin@ms2.inr.ac.ru}
\abstract{We consider oscillons~--- localized, quasiperiodic, and
  extremely long-living classical solutions in models with real
  scalar fields. We develop their effective description in the limit of
  large size at finite field strength. Namely, we note that nonlinear
  long-range field configurations can be described by an effective complex
  field $\psi(t, \boldsymbol{x})$ which is related to the original
  fields by a canonical 
  transformation. The action for $\psi$ has the form of a systematic
  gradient expansion. At every order of the expansion, such an effective 
  theory has a global U(1) symmetry and hence a family of stationary
  nontopological solitons~--- oscillons. The decay of the latter objects is a
  nonperturbative process from the viewpoint of the effective 
  theory.   Our approach gives an intuitive understanding of oscillons
  in full nonlinearity and  explains their longevity. Importantly, it
  also provides reliable selection criteria for models with long-lived 
  oscillons. This technique is more precise in the nonrelativistic
  limit, in the notable cases of nonlinear,  extremely long-lived, and
  large objects, and also in lower spatial dimensions. We test the
  effective theory by performing explicit numerical simulations of a
  $(d+1)$-dimensional scalar field with a plateau potential. 
} 

\begin{document}
\preprint{INR-TH-2022-017}

\maketitle
\section{Introduction}
\label{sec:intro}
Oscillons~\cite{Gleiser:1993pt} are compact, almost periodic, and
long-lived classical solutions in models with real bosonic fields,
notably, the scalar field $\varphi(t,\, \boldsymbol{x})$. These objects 
were discovered in explicit numerical simulations in the
70s~\cite{Kudryavtsev:1975dj, Bogolyubsky:1976nx}, but even now 
theoretical reasons for their widespread existence and extreme
longevity are poorly understood. To date, the oscillons were found
in a plethora of  theories with attractive
self-interactions~\cite{Kolb:1993zz, Piette:1997hf, 
  Gleiser:2008dt, Amin:2010jq, Gleiser:2010qt, Salmi:2012ta,
  Amin:2013ika, Sakstein:2018pfd, Olle:2019kbo, 
  Zhang:2021xxa}. All of them radiate waves and eventually disappear, but
before that live for $10^3$ oscillation
cycles in generic models~\cite{Zhang:2020bec} and up to
$10^{14}$ cycles in special cases~\cite{Olle:2020qqy}. Such large
numbers are highly remarkable and deserve to be explained, as the
models with oscillons usually lack large, small, or fine-tuned parameters.

Meanwhile, the oscillons are becoming a workhorse in cosmology. They
nucleate excessively during generation of axion~\cite{Kolb:1993hw,
  Vaquero:2018tib, Buschmann:2019icd, Gorghetto:2020qws} or
ultra-light~\cite{OHare:2021zrq} dark matter, may accompany
cosmological phase transitions~\cite{Gleiser:1993pt, Copeland:1995fq,
  Dymnikova:2000dy, Farhi:2007wj, Gleiser:2010qt, Bond:2015zfa} and be formed
by the oscillating inflaton field during preheating~\cite{Amin:2010dc,
  Amin:2011hj,Hong:2017ooe, Sang:2020kpd}. Their close relatives --- gravitationally 
bound boson stars --- appear in the  centers of the smallest
axion~\cite{Levkov:2018kau, Eggemeier:2019jsu, Chen:2020cef, Chan:2022bkz} and
fuzzy~\cite{Schive:2014dra, Schive:2014hza, Veltmaat:2018dfz} dark
matter structures. Cosmological oscillons may produce gravitational
waves~\cite{Zhou:2013tsa, Liu:2017hua, Lozanov:2019ylm, Sang:2019ndv}, participate
in baryogenesis~\cite{Lozanov:2014zfa}, conceive axion
miniclusters~\cite{Kolb:1993hw, Vaquero:2018tib, Xiao:2021nkb}, or
create a population of primordial black holes~\cite{Cotner:2019ykd,
  Kou:2019bbc}, cf.~\cite{Garani:2021gvc}. With sufficiently large
lifetimes, they may even act as  
dark matter candidates~\cite{Olle:2020qqy}. All of this
requires more systematic studies of these fascinating objects. 

Presently, the only model-independent method to describe oscillons
is based on nonrelativistic (small-amplitude)
expansion~\cite{Dashen:1975hd, Kosevich1975, Fodor:2008es,
  Fodor:2019ftc}. This technique applies to quasiperiodic 
field configurations with sufficiently large sizes~$R$, weak fields
$\varphi(t,\, \boldsymbol{x})$, and oscillation frequencies~$\omega$
nearing the field mass~$m$, 
\begin{equation}
  \label{eq:1}
  \mbox{small-amplitude:} \qquad  R \gg m^{-1}\;, \qquad \omega \approx m\;, \qquad
  \varphi\; \mbox{ is small}\;.
\end{equation}
In terms of particle physics, such configurations describe  condensates
of weakly interacting nonrelativistic bosons with small
momenta~$R^{-1}$ and low binding energies~${m-\omega \ll m}$. Since
the particle number $N$ is conserved in the nonrelativistic limit, it
is natural to expect that the bosons form stable localized lumps~---
oscillons~--- if their self-interactions are attractive. Solving  the 
classical field equations  order-by-order in the field amplitude, one
can compute the profiles and energies  of 
oscillons~\cite{Dashen:1975hd, Kosevich1975,
  Fodor:2008es}. Besides, exponentially small effects complementary to
the nonrelativistic expansion describe radiation from these objects and
estimate their lifetimes~\cite{Segur:1987mg,
  Fodor:2008du,Fodor:2009kf, Fodor:2019ftc}.  

The above technique is unsatisfactory in three respects. First, it
fails to describe exceptionally long-lived, large-amplitude, and large-size
oscillons~\cite{Olle:2019kbo, Zhang:2020bec, Olle:2020qqy} discovered
in scalar models with almost quadratic monodromy
potentials~\cite{Silverstein:2008sg, McAllister:2008hb}. In
some other cases, the nonrelativistic expansion requires a
modification~\cite{Amin:2010jq}. To explain stability of
large-amplitude  oscillons,
Refs.~\cite{Kasuya:2002zs, Kawasaki:2015vga} 
suggested existence of  an adiabatic invariant that is approximately
conserved during evolution of nonlinear oscillating fields. This
quantity generalizes the particle number
$N$.  Then the oscillons minimize the energy at a given value of the
invariant, similarly to  Q-balls~\cite{Friedberg:1976me,
  Coleman:1985ki, Nugaev:2019vru}. However, a consistent off-shell and
strong-field 
definition of the adiabatic invariant is absent so far.

Second, the above
nonrelativistic expansion is asymptotic. In some models, it poorly
approximates oscillon profiles even if the value of
the nonrelativistic parameter is small~\cite{Fodor:2008es}. Third, generic
three-dimensional oscillons with $\omega\approx m$ are, in fact,
unstable~\cite{Zakharov12}. In this  case the lowest term of
the small-amplitude expansion gives qualitatively incorrect
predictions for oscillons with lower  $\omega$.

In this paper we develop an Effective Field Theory (EFT) description
of oscillons with large size and any amplitude,
\begin{equation}
  \label{eq:2}
  \mbox{EFT:} \qquad R \gg m^{-1}\;, \qquad \mbox{$\varphi$ is arbitrary.}
\end{equation}
This regime is natural: crude estimates
show~\cite{Gleiser:2004an} that long-living oscillating objects can exist 
only in low enough dimensions and with large enough sizes. For
definiteness we will consider one real scalar field
$\varphi$ with symmetric potential; generalization to other cases is
straightforward. 

We observe that as long as the spatial scales are large and the
gradient terms in the equations are suppressed, the classical field
$\varphi(t,\, \boldsymbol{x})$ at any $\boldsymbol{x}$ performs fast,
almost mechanical oscillations in the scalar potential
$V(\varphi)$. Then the proper slowly varying EFT
variables are the  amplitude $I(t,\, \boldsymbol{x})$ and phase
$\theta(t,\, \boldsymbol{x})$ of the oscillations~--- the action 
and angle variables in the mechanical system with potential
$V$. These quantities can be combined into one complex EFT field,
\begin{equation}
  \label{eq:3}
  \psi(t,\, \boldsymbol{x}) = \sqrt{I} \, \cdot \mathrm{e}^{-i\theta}\;.
\end{equation}
From the viewpoint of the original theory, $\psi$ and $\psi^*$ are related to
$\varphi$ and $\partial_t \varphi$ by a canonical transformation. We
demonstrate that the effective classical action for their slowly-varying parts
can be written in the form of a systematic gradient expansion, where 
the terms like  $|\partial_{\boldsymbol{x}} \psi|^2$ and
$|\partial_{\boldsymbol{x}} \psi|^4$ appear in the first and
second orders, respectively. Most importantly, at any order the
effective action is invariant under global U(1) symmetry~$\psi \to
\psi\, \mathrm{e}^{-i\alpha}$ and hence has a conserved charge 
\begin{equation}
  \label{eq:4}
  N = \int d^d \boldsymbol{x} \, |\psi|^2 + \mbox{corrections}\;.
\end{equation}
When evaluated in the leading order on the classical solution, this
charge coincides with the adiabatic 
invariant of Refs.~\cite{Kasuya:2002zs, Kawasaki:2015vga}. We conclude
that our effective theory possesses a family of oscillons which are
indeed the nontopological solitons minimizing the energy at a
given~$N$.  

Notably,  the same effective approach may be useful
  for studying some mechanical systems with many degrees of
  freedom, see Appendix~\ref{sec:Mech_Example} for details. 

Our classical EFT is way simpler than direct
computation of tree diagrams suggested in~\cite{Braaten:2015eeu,
  Braaten:2016kzc, Visinelli:2017ooc}  and more powerful than their
partial resummation in~\cite{Eby:2014fya}. It generalizes
nonrelativistic EFT approaches of~\cite{Mukaida:2016hwd,
  Eby:2018ufi, Salehian:2021khb} which can be restored by re-expanding
the effective action in field amplitudes. The latter re-expansion also
reproduces all small-amplitude results~\cite{Dashen:1975hd,
  Kosevich1975, Fodor:2008es, Fodor:2019ftc} for oscillons. 

But even more importantly, the effective theory can explain essentially nonlinear
large-amplitude oscillons~\cite{Amin:2010jq, Olle:2019kbo,
  Zhang:2020bec, Olle:2020qqy} and clarify conditions for their
existence and longevity. Namely, these objects exist within
the EFT if the scalar potential of the model satisfies
certain requirements summarized in Sec.~\ref{sec:conditions-existence}
and in Discussion. For linear stability, their charges $N(\omega)$
should satisfy  the
Vakhitov-Kolokolov criterion~\cite{vk, Zakharov12, Amin:2010jq,
  Nugaev:2019vru}. The decay of oscillons is expected to be
a nonperturbative process from the viewpoint of the gradient expansion, 
similarly to the nonrelativistic case~\cite{Segur:1987mg,
  Fodor:2008du, Fodor:2009kf,  Fodor:2019ftc}. We do not consider such
processes in this paper but note that oscillons should be 
exponentially long-lived whenever Eq.~(\ref{eq:2}) holds and the
EFT works. This leads to additional conditions on the potential, see  
Sec.~\ref{sec:long-stab-oscill} and Discussion. 

\begin{sloppy}

We test the EFT by performing spherically symmetric
simulations of a $(d+1)$-dimensional scalar field with the plateau
potential which is typical  
for $\alpha\mbox{-attractor}$ inflation~\cite{Kallosh:2013hoa} and
is frequently employed in oscillon studies~\cite{Zhang:2020bec}. We
find that the EFT correctly describes oscillons:
qualitatively in~$d=3$ dimensions, precisely in $d=1$ and~2, and
exactly in the nonrelativistic limit~${\omega \to m}$. 

\end{sloppy}

As a final result, we explain significant dependence of the oscillon
stability properties on the dimensionality $d$ of
space~\cite{Gleiser:2004an}. Note that full classical equation for  
the  spherically-symmetric scalar field $\varphi(t,\, r)$ has a nontrivial formal
limit $d\to 0$. We observe that in $d=0$ it has an ever-lasting and
strictly periodic stationary solution~--- a ``zero-dimensional
oscillon.'' This explains why  oscillons mostly appear in
low-dimensional models and become increasingly
unstable~\cite{Gleiser:2004an} at larger $d$. Amusingly,  
our EFT also becomes exact in $d\to 0$, since all its
higher-order terms are proportional to $d$. As a consequence, it
works better in lower dimensions.

The paper is organized as follows. We start  with  an explicit
numerical example of oscillon in
Sec.~\ref{sec:oscill-numer-ilustr}. Then we introduce the leading-order 
EFT in Sec.~\ref{sec:Low-energy-Theory} and illustrate it in a
mechanical model in Appendix~\ref{sec:Mech_Example}. In 
Sec.~\ref{sec:oscill-effect-theory}, we study solitonic oscillons
within the EFT and formulate the conditions for their existence and
longevity. Section~\ref{sec:Num_Tests} tests  effective theory against
explicit numerical simulations. The formal limit of zero
dimensions and higher-order corrections are considered in
Secs.~\ref{sec:limit-d-to-zero} and~\ref{sec:Corrections},
respectively. For clarity, we calculate the corrections in the mechanical model of Appendix~\ref{sec:Mech_Example} as well. 
We perform comparison with small-amplitude expansion
in Sec~\ref{sec:limit-omega-to-zero} and discuss future prospects  in
Sec.~\ref{sec:conclusions}. Appendices \ref{Appendix:Num} and 
\ref{sec:one-dimens-oscill} to~\ref{Append:Corrections_F} include 
details of numerical and analytical calculations.

%%%%%%%%%%%%%%%%%%%%%%%%%%%%%%%%%%%%%%%%%%%%
\section{Oscillons: numerical illustration}
\label{sec:oscill-numer-ilustr}
We consider real scalar field $\varphi(t,\, \boldsymbol{x})$ with
nonlinear potential $V(\varphi)$ in $(d+1)$ dimensions. It satisfies 
the equation 
\begin{equation}
  \label{eq:5}
  (\partial_t^2 -\Delta) \varphi = - V'(\varphi)\;,
\end{equation}
where  $\Delta$ is a $d$-dimensional Laplacian and the prime denotes
$\varphi$ derivative. Let us numerically 
demonstrate that the oscillons appear if $V(\varphi)$ is chosen appropriately.

For these purposes, we select $d=3$ dimensions, a potential motivated
by $\alpha$-attractor inflation~\cite{Kallosh:2013hoa,
  Zhang:2020bec}, 
\begin{equation}
  \label{U_def}
  V(\varphi) = \frac{1}{2} \tanh^2 \varphi\;,
\end{equation}
and adopt dimensionless units\footnote{Introduced by rescaling
  ${t \to  t/m}$, $\boldsymbol{x} \to
  \boldsymbol{x}/m$, and $\varphi \to \Lambda \varphi$ in the model with
  canonically normalized field and potential ${V = \frac12 m^2\Lambda^2  \,
    \mathrm{tanh}^2(\varphi/\Lambda)}$.} with field mass equal to one,
$m=1$. Importantly, the potential~(\ref{U_def}) is attractive,
i.e.\ grows slower than $\varphi^2$.  This property is
usually held responsible for formation of long-living lumps~---
oscillons.

Starting from the Gaussian initial data  $\varphi = \varphi_0 \exp
(-r^2 / \sigma^2)$ and $\partial_t \varphi =0$ at~${t=0}$, we 
numerically solve equation for the spherically-symmetric field $\varphi(t,\, 
r)$, where $r \equiv |\boldsymbol{x}|$ is the radial
coordinate. Details\footnote{In short, we employ an infinite-order spatial
  discretization based on fast Fourier transform (FFT), absorb the outgoing
  radiation with the artificial
  damping~\cite{Gleiser:1999tj}, and use 
  fourth-order symplectic Runge-Kutta-Nystr\"om integrator.} of
this procedure are presented in Appendix~\ref{Appendix:Num}. 
\begin{figure}[h]
  \centerline{\includegraphics{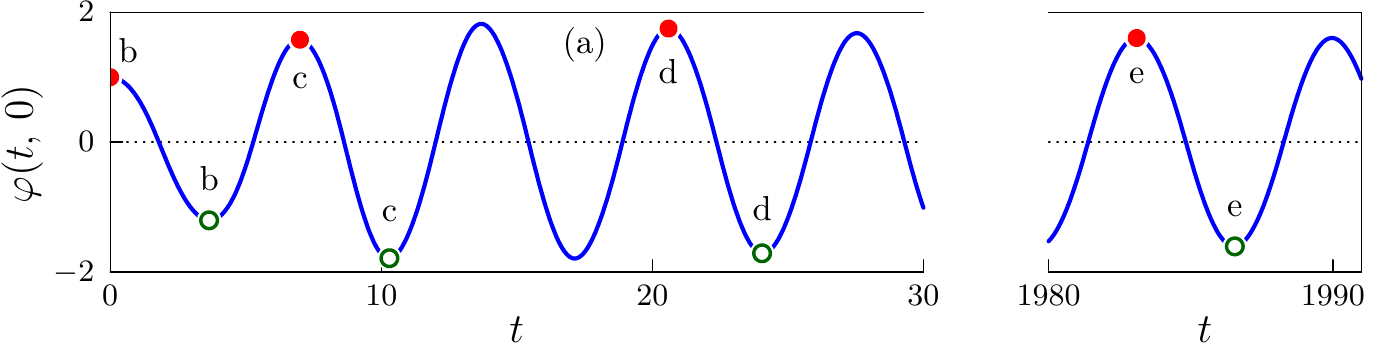}}
  \centerline{\includegraphics{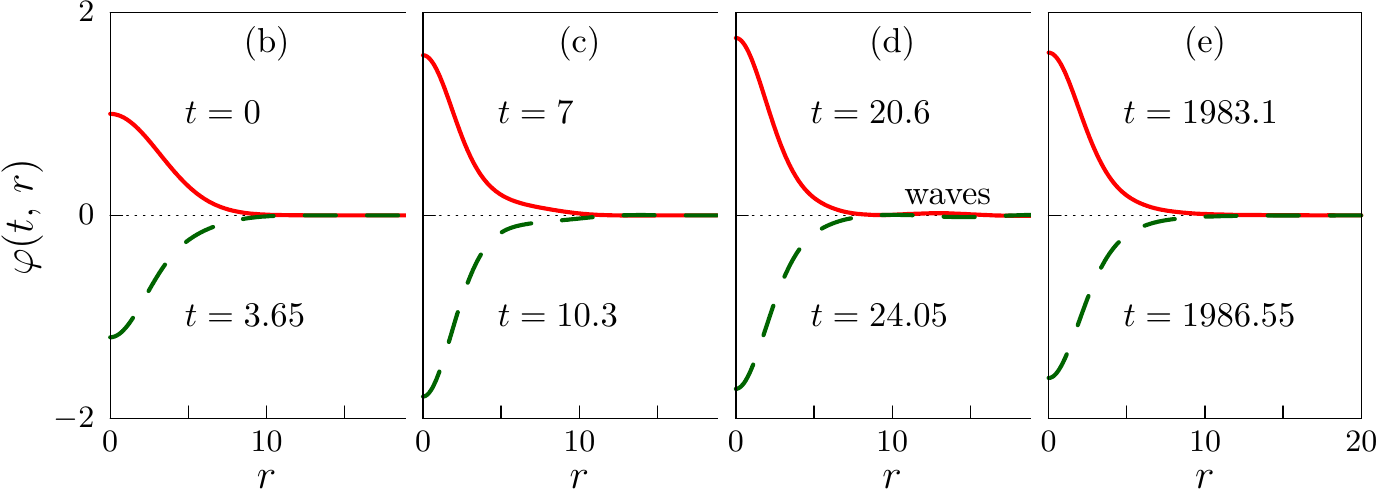}}
  \caption{Numerical evolution of a spherically-symmetric scalar field
    $\varphi(t,\, r)$ with Gaussian initial data ($\varphi_0 =
    1$ and $\sigma = 20$) in the $(3+1)$-dimensional
    model (\ref{U_def}). Figure (a) shows the field~$\varphi(t,\,0)$
    in the center~${r=0}$ as a function of  time, while Figs.~(b)---(e) 
    display configurations at time moments of maximal (solid
    lines)  and minimal (dashed lines) values of $\varphi(t,\,
    0)$. The latter are marked in Fig.~(a) with filled and empty
    circles, respectively.\label{fig:3d_evolution}} 
\end{figure}
The result  is demonstrated in Fig.~\ref{fig:3d_evolution} and in the 
movie~\cite{movie}. After a  
short period of aperiodic nonlinear wobbling
(Figs.~\ref{fig:3d_evolution}b,c and the left-hand side of
Fig.~\ref{fig:3d_evolution}a) the 
field lump in the center shakes off some outgoing waves
(tiny, but seen in Fig.~\ref{fig:3d_evolution}d) and settles in an almost periodic  
long-living configuration (Fig.~\ref{fig:3d_evolution}e and the
right-hand side of Fig.~\ref{fig:3d_evolution}a).  This is the oscillon to
be studied in what follows.

We stress that oscillons are not specific to the model~(\ref{U_def}) or
to the Gaussian initial data. They appear in the spectacular number of
attractive theories in low enough~$d$,  see Refs.~\cite{Kolb:1993zz,
  Piette:1997hf, Gleiser:2008dt, Amin:2010jq,
  Gleiser:2010qt, Salmi:2012ta, Amin:2013ika, Sakstein:2018pfd,
  Olle:2019kbo, Zhang:2020bec, Cyncynates:2021rtf, Zhang:2021xxa}.  Although the
model~(\ref{U_def}) is generic from the viewpoint of oscillon
longevity, these objects typically survive to the end of our simulations
lasting up to $10^{5}$ cycles. Such lifetimes deserve to be explained.

%%%%%%%%%%%%%%%%%%%%%%%%%%%%%%%%%%%%%%%%%%%%
\section{Classical EFT}
\label{sec:Low-energy-Theory}
Now, we construct  classical effective field theory (EFT) for
nonlinear oscillons in the limit of large size (\ref{eq:2}) or, more
specifically, at
\begin{equation}
  \label{eq:6}
  |\partial_{i} \varphi| \ll m \varphi\;,
\end{equation}
where the field mass is restored and $\partial_i$ is the spatial
derivative. Soon we will see that in generic 
models this condition gets parametrically satisfied only in the
nonrelativistic limit~\cite{Fodor:2008es} when the oscillon
frequency~$\omega$ approaches $m$ and the field amplitude becomes
small. But there are also special models~\cite{Olle:2019kbo,
  Zhang:2020bec, 
  Olle:2020qqy} with exceptionally
long-lived and large-amplitude oscillons, the sizes of which are
proportional to large parameters. To 
cover both cases, we work at  finite frequency~$\omega$ and
consider nonlinear fields.

\begin{sloppy}

Imagine that in the roughest approximation we can ignore the 
term with the spatial derivatives in the field equation~(\ref{eq:5}). This
leaves a nonlinear mechanical system~${\partial^2_{t} \varphi = -
  V'(\varphi)}$ which can be solved in the action-angle
variables~\cite{LL1, Arnold}. Namely, one introduces the
momentum $\pi_\varphi = \partial_t \varphi = \sqrt{2h   -
  2V(\varphi)}$, where $h$ is a mechanical energy, and then performs
a canonical transformation to the new variables $I$ and~$\theta$,  
\begin{equation}
  \label{eq:7}
  \varphi = \Phi(I, \theta)\;, \qquad\qquad  \partial_t \varphi  =
  \pi_\varphi = \Pi (I, \theta)\;,
\end{equation}
in such a way that the action $I$ is conserved during 
the mechanical motion and the angle~$\theta$ is canonically conjugate to
it. More explicitly, 
\begin{equation} 
  \label{I(h)}
  I(h) = \frac{1}{2\pi} \oint \pi_\varphi (h,\, \varphi)  \, d\varphi \;, 
\end{equation}
where the integration is done over the oscillation period, and the
angle equals
\begin{equation}
  \label{theta_def}
  \theta(I,\, \varphi) = \frac{\partial}{\partial I} \; \int_{\varphi_h}^\varphi
  \pi_\varphi (h(I),\, \varphi') \, d\varphi' \;.
\end{equation}
Hereafter, $h(I)$ is obtained by inverting Eq.~(\ref{I(h)}) and
$\theta=0$ corresponds to a turning point $\varphi = 
\varphi_h$ with~$\pi_\varphi = 0$. Recalling
that~${h = \pi_\varphi^2 / 2 + V(\varphi)}$, one can express $\varphi$ and
$\pi_\varphi$ from Eqs.~(\ref{I(h)}) and~(\ref{theta_def})  
obtaining~$\Phi(I,\, \theta)$ and $\Pi (I,\,  
\theta)$. Note that the latter functions can be evaluated explicitly for
some potentials,  approximately in other cases, or efficiently
represented in the form of convergent power series using computer
algebra. Once this is done, the solution to the mechanical equation
would be~${I   =  \mbox{const}}$ and $\theta =\Omega \,  t +
\mbox{const}$, where $\Omega(I) = dh/dI$ is the frequency of
oscillations in the potential $V(\varphi)$. As designed, $I$ remains
constant during the nonlinear oscillations and $\theta$ increases by $2\pi$
every period.

\end{sloppy}

In field theory, the terms with the spatial derivatives cannot be ignored
altogether even if they are suppressed, as they 
determine the spatial profile of  the oscillon. Nevertheless,
Eq.~(\ref{eq:7}) still defines a 
canonical transformation from $\varphi$ and $\pi_\varphi = \partial_t \varphi$
to $I(t, \,\boldsymbol{x})$ and~$\theta(t,\, \boldsymbol{x})$. The
latter variables are not the true action and angle in field theory, but
they still parameterize the amplitude and phase of local 
oscillations at every point. It is natural to expect  that $I$ and
$\partial_t\theta$ slowly depend on space and time in  
the limit of large size, unlike the fast-oscillating~${\varphi(t,\,
  \mathbf{x})}$. We will use them as smooth variables in the 
leading-order EFT.  

Now, we evaluate classical effective action for slowly varying
$I$ and $\theta$. We substitute the
transformation~(\ref{eq:7}) into the action of a scalar field, 
\begin{equation} 
  \label{eq:9}
  {\cal S} = \int dt \, d^d \boldsymbol{x} \left( \pi_\varphi \partial_t \varphi  -
  h - \frac12(\partial_i \varphi)^2
  \right)\;, \qquad h \equiv \pi_\varphi^2/2 + V(\varphi)\;,
\end{equation}
where the ``mechanical Hamiltonian'' $h(\varphi,\, \pi_\varphi)$ is
introduced for convenience. Since the transformation~(\ref{eq:7}) is 
canonical, $\int dt \, \pi_\varphi \partial_t \varphi = \int dt
\, I \partial_t \theta$. Besides, $h$ is a function of $I$ defined 
in Eq.~(\ref{I(h)}). Finally, we note that the subdominant term with spatial
derivatives~${f = (\partial_i \varphi)^2}$ is integrated in the
action~\eqref{eq:9} over many oscillation periods. Thus, let us
explicitly average it over period, i.e.\ over $\theta$ which changes
almost linearly in time,
\begin{equation}
  \label{eq:10}
  \langle f \rangle = \frac{1}{2\pi} \int_{0}^{2\pi} 
  f(I, \theta) \,d\theta \;.
\end{equation}
Using Eq.~(\ref{eq:7}), we find,  
\begin{equation}
  \label{mu_coeff}
  \langle (\partial_i \varphi )^2 \rangle  \approx \frac{(\partial_i
    I)^2}{\mu_I(I)} + \frac{(\partial_i \theta)^2}{\mu_\theta(I)}\;, \qquad 
  \qquad \frac{1}{\mu_I} \equiv \left\langle
  (\partial_I \Phi)^2 \right\rangle \;, \qquad
    \frac{1}{\mu_\theta} \equiv \left\langle 
  (\partial_\theta \Phi)^2 \right\rangle\;,
\end{equation}
where we moved all slowly varying quantities $\partial_i I$ and
$\partial_i \theta$ out of the averages, introduced~${\partial_I \equiv
\partial / \partial I}$ and $\partial_\theta \equiv \partial / \partial 
\theta$, and observed that the cross-term  $\langle\partial_I
  \Phi \,\partial_\theta  \Phi \rangle  \partial_i I \partial_i \theta$ 
vanishes due to time reflection symmetry $t \to -t$, $\theta \to
-\theta$. Indeed, recall that $\theta=0$ corresponds to a turning
point $\pi_\varphi=0$. Then $\Phi$ is an
even function of $\theta$ and~$\partial_I   \Phi \partial_\theta \Phi$
is odd with zero average. 

Collecting the terms, we obtain the leading-order effective action,
\begin{equation}
  \label{action_eff_I}
  \mathcal{S}_\mathrm{eff}  = \int dt \, d^d {\boldsymbol{x}} \left(I \partial_t
    \theta - h(I) - \frac{(\partial_iI)^2}{2 \mu_I(I)} -
    \frac{(\partial_i\theta)^2}{2 \mu_\theta(I)} \right)\;,
\end{equation}
where $\mu_I$ and $\mu_\theta$ are explicitly given by 
Eqs.~(\ref{mu_coeff}) if the transformation $\Phi$ is
known. The same action takes more familiar form in terms of a complex
field $\psi(t,\, \boldsymbol{x})$ introduced in Eq.~(\ref{eq:3}),
\begin{equation}
  \label{action_eff_psi}
  \mathcal{S}_\mathrm{eff} = \int dt \; d^d \boldsymbol{x} \left( i \psi^*
    \partial_t \psi - h(|\psi|^2) - \frac{|\partial_i \psi|^2}{2
      \mu_1} - \frac{1}{2\mu_2} \left[\psi^{*2} (\partial_i \psi)^2 +
    \mathrm{h.c.}\right] \right)\;,
\end{equation}
where $\mu_1 \equiv 2I \mu_I \mu_\theta / (4I^2 \mu_\theta + \mu_I)$
and $\mu_2 \equiv 4I^2 \mu_I \mu_\theta / (4I^2 \mu_\theta -
\mu_I)$. We arrived at a nonlinear Schr\"odinger model for $\psi$ with
the form factors $h$, $\mu_1$, and $\mu_2$ depending on $I
\equiv |\psi|^2$. An equation for the evolution of
long-range fields can be obtained by varying Eq.~(\ref{action_eff_I})
over $I$ and $\theta$, or Eq.~(\ref{action_eff_psi}) over~$\psi$
and~$\psi^*$. In particular,  
\begin{equation}
  \label{eq:11}
  \partial_t I = \partial_i (\partial_i
  \theta / \mu_\theta )\;, \qquad\qquad  \partial_t \theta =
  \Omega - \frac{\Delta I}{\mu_I} + \frac{\partial_I \mu_I}{2\mu_I^2}
  \, (\partial_i I)^2- \frac{\partial_I \mu_{\theta}}{2\mu_\theta^2} 
  \, (\partial_i \theta)^2\;, 
\end{equation}
with $\Omega \equiv  \partial_I h$. An equivalent
equation for $\psi$ has the form of a modified nonlinear Schr\"odinger
equation. 
 
It is worth stressing that the above effective approach is approximate
due to period averaging in Eq.~\eqref{mu_coeff}. We will see
in Sec.~\ref{sec:Corrections} that corrections to the effective action 
are suppressed by at least four spatial derivatives of $I$ and
$\theta$. This will confirm that the effective theory works for
the long-range fields, indeed. 

To support the classical EFT even further, we demonstrate in
Appendix~\ref{sec:Mech_Example} that it correctly reproduces the
spectrum of two weakly coupled mechanical oscillators. We  expect
that in the future this method may be helpful for studies of some
nonlinear dynamical systems.

The most important property of the effective theory is a global
U(1) symmetry
\begin{equation*}
  \theta \to \theta + \alpha \qquad\mbox{or} \qquad \psi \to \psi \, 
  \mathrm{e}^{-i\alpha}\;,
  \quad \psi^* \to \psi^* \, \mathrm{e}^{i\alpha}\;,
\end{equation*}
which appears after averaging over $\theta$ in Eq.~(\ref{mu_coeff}). As a
consequence, the U(1) charge $N$ given by the first term in
Eq.~(\ref{eq:4}) conserves:  $\partial_ t N = 0$ according to
Eqs.~(\ref{eq:11}). On-shell,
the value of  this charge coincides with the ``adiabatic
invariant'' in Refs.~\cite{Kasuya:2002zs, Kawasaki:2015vga}:  
\begin{equation}
   \label{eq:8}
  N = \frac{1}{2\pi} \int d^d \boldsymbol{x}\oint \pi_\varphi \, d\varphi =
  \frac{1}{2\pi} \int d^d
  \boldsymbol{x} \oint dt \,  (\partial_t \varphi)^2 
\end{equation}
where all integrals cover for one oscillation period and we used
Eq.~(\ref{I(h)}). Note that Eq.~\eqref{eq:8} is convenient for
finding the values of $N$ on quasiperiodic solutions, but unlike
Eq.~(\ref{eq:4}), it is useless off-shell.
In the next Section we will observe that under certain conditions 
conservation of charge leads to appearance of stable nontopological
solitons\footnote{Called ``$I$-balls''   in~\cite{Kasuya:2002zs}.}
similar to $Q$-balls~--- the oscillons.

\begin{sloppy}

We finish this Section by illustrating the calculation of the
leading-order effective action in the
model~(\ref{U_def}). Equation~(\ref{I(h)}) gives\footnote{Change of
  variables to $\beta = \tanh\varphi \,\sqrt{1-2h}   \, / \sqrt{2h
    -\tanh^2 \varphi}$ turns Eqs.~(\ref{I(h)}), (\ref{theta_def}) into
  rational integrals.},
\begin{equation}
  \label{eq:12}
  h(I) = I - I^2 / 2\;.
\end{equation}
This means that the frequency of mechanical oscillations in the
potential $V(\varphi)$ is ${\Omega = \partial_I h = 1 - I}$.
The canonical transformation $\varphi = \Phi (I, \, \theta)$
and~$\pi_\varphi = \Pi (I,\, \theta)$ is then obtained using
Eq.~(\ref{theta_def}) and the definition of $h (\varphi,\, \pi_\varphi)$:
\begin{equation}
  \label{tanh_action_angle}
  \Phi = \operatorname{arcsinh}\left(\frac{\sqrt{I(2-I)}}{1-I}\, \cos
  \theta\right), \qquad \Pi = - \frac{(1-I)\sin
    \theta}{\sqrt{[I(2-I)]^{-1}-\sin^2 \theta}}\;.
\end{equation}
One can readily check that the Poisson bracket of these functions
equals one. Next, we evaluate\footnote{\label{fn:1}Due to time reflection symmetry,
  the integrands in Eq.~(\ref{mu_coeff}) are symmetric
  functions,~${f(\theta) = f(-\theta)}$. In this 
  case the averaging~(\ref{eq:10}) is given by the 
  contour integral ${\langle f \rangle =  \oint dz \,  f (z) / (2\pi i
    z)}$ along the unit circle~${z = -\mathrm{e}^{2i\theta}}$, $|z| =
  1$, which can be computed by residuals.} the integrals~(\ref{mu_coeff}) over
$\theta$ and obtain the  time-averaged form factors,   
\begin{equation}
  \label{mu_coeff_th}
  \mu_I = I(2-I)^2 (1-I)^2\;, \qquad\qquad
  \mu_\theta = I^{-1}-1\;.
\end{equation}
The coefficients in the action for $\psi$ immediately follow: 
\begin{equation}
  \label{eq:13}
  \mu_1 = \frac{2(1-I)^2 (2-I)^2}{4 + (1-I)(2-I)^2}\;, \qquad \qquad
  \mu_2 = \frac{4I(1-I)^2 (2-I)^2}{4 - (1-I)(2-I)^2}\;;
\end{equation}
notably, $\mu_{1}$ and $\mu_2 / I$ are finite in the weak-field
limit $I \to 0$. We conclude that once the scalar
potential is fixed, the EFT has an explicit form of a nonlinear
Schr\"odinger-like model for $\psi$. 

\end{sloppy}

%%%%%%%%%%%%%%%%%%%%%%%%%%%%%%%%%%%%%%%%%%%
\section{Oscillons in the effective theory}
\label{sec:oscill-effect-theory}

\subsection{Conditions for existence}
\label{sec:conditions-existence}
In certain cases, the  EFT has a family of compact
nontopological solitons~--- oscillons. Indeed, the  stationary Ansatz 
\begin{equation}
  \label{eq:14}
  \psi = \psi (\boldsymbol{x}) \, \mathrm{e}^{-i\omega t} \,, \;\;
  \psi(\boldsymbol{x}) \in \mathbb{R}\,, \qquad \qquad \mbox{or}
  \qquad \qquad I = \psi^2 (\boldsymbol{x})\;, \qquad \theta = \omega t
\end{equation}
passes Eqs.~(\ref{eq:11}) and gives a  
  Schr\"odinger-like equation for the real-valued 
oscillon profile~$\psi (\boldsymbol{x})$,
\begin{equation} 
  \label{i_ball_profile_generic}
 -  \frac{2 \psi^2}{\mu_I}\, \Delta \psi - 
 (\partial_i \psi)^2 \frac{d}{d \psi} \left(
    \psi^2/\mu_I\right)  +
 \Omega\psi = \omega \psi \;,
\end{equation}
where the ``mass'' $\mu_I / \psi^2$ and ``potential'' $\Omega =
\partial_I h$
are the functions of $I \equiv \psi^2(\boldsymbol{x})$. Oscillons exist
whenever Eq.~\eqref{i_ball_profile_generic} has stable localized solutions
at some $\omega$.

Let us deduce general requirements for existence of
oscillons. Our analysis will be based  on conservation
laws~\cite{Coleman:1985ki, Nugaev:2019vru}. Observe that 
the oscillon configurations minimize the  energy at a fixed 
charge $N$, i.e.\ extremize the functional 
\begin{equation}  
  \label{F_generic}
  F = E - \omega N = \int d^d \boldsymbol{x} \left[h(I) +
    \frac{(\partial_iI)^2}{2 \mu_I(I)} + \frac{(\partial_i\theta)^2}{2
      \mu_\theta(I)} - \omega I \right]\;,
\end{equation}
where $\omega$ is a Lagrange multiplier. Indeed, minimization
over $\theta$ gives ${\partial_i\theta = 0}$, after which $(-F)$
coincides with the Lagrangian in Eq.~\eqref{action_eff_I} evaluated on
the configuration with $\theta=\omega t$. Thus,  the
profile equation~(\ref{i_ball_profile_generic}) can be obtained by
extremizing $F$ over~${I =\psi^2 (\boldsymbol{x})}$. We immediately
see the interpretation of the oscillon frequency~$\omega$. Since~$F$
is extremal with respect to all fields, $\delta E = \omega\, \delta N$
under any variation. In particular, small shift of $\omega$ gives,
\begin{equation}
  \label{E_N_Om_relations}
  \omega = dE/dN  \qquad \mbox{and} \qquad d
    F / d\omega = - N\;,
\end{equation}
where Eq.~(\ref{F_generic}) was differentiated in the second equality. If $N$ is
the oscillon charge, $\omega$ is the chemical potential~--- energy per unit charge,
and $F$ is the grand thermodynamic potential of the entire object. 

It is instructive to introduce the field
  \begin{equation}
    \label{eq:26}
\chi(I) = \int_0^{I} dI' \, [\mu_I(I')]^{-1/2} = 2\int_0^{\psi}
\psi'd\psi' /  \sqrt{\mu_I}\;,
\end{equation}
with canonically normalized gradient term:
\begin{equation}
  \label{F_remaining}
  F = \int d^d \boldsymbol{x} \left[ \frac12 (\partial_i \chi)^2 -
    U_{\omega}(\chi) \right]\;, \qquad \mbox{where} \qquad 
  \qquad U_{\omega}
  \equiv \omega I - h\;.
\end{equation}
\setlength{\columnsep}{.5cm}
\begin{wrapfigure}[13]{r}{5,7cm}
  \centering
  \includegraphics{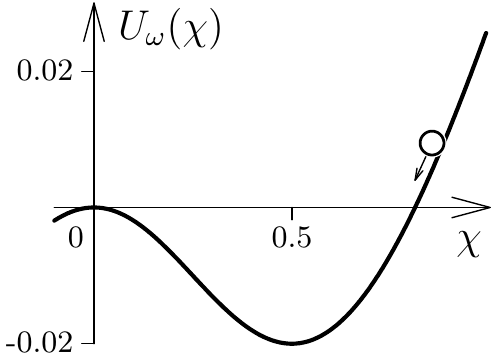}
  \caption{Mechanical potential $U_\omega(\chi)$ plotted for $\omega =
    0.8$  in the model~(\ref{U_def}).}
  \label{fig:mech_pot}
\end{wrapfigure}
Now, we can study the oscillons with methods 
developed for $Q$-balls~\cite{Coleman:1985ki, Nugaev:2019vru}. We will
assume that they are spherically symmetric ${\chi =   \chi(r)}$:
numerical simulations indeed indicate that angular asymmetry
disappears immediately after 
formation of these objects~\cite{Adib:2002ff}, just  
like in the case of gravitationally bound Bose
stars~\cite{Dmitriev:2021utv}. The profile equation for $\chi(r)$
takes the form, 
\begin{equation}
  \label{eq:16}
  \partial_r^2 \chi + \frac{d-1}{r} \,\partial_r \,\chi  = - dU_\omega / d\chi\;.
\end{equation}
Notably, Eq.~(\ref{eq:16}) coincides with the Newton's law for a unit
mass particle moving with ``time'' $r$ in the mechanical  potential
$U_\omega(\chi)$, see Fig.~\ref{fig:mech_pot}. The second term in
Eq.~(\ref{eq:16}) describes friction that decreases the mechanical energy in
$d>1$. Since the oscillon is regular and localized, we impose boundary 
conditions $\partial_r\chi = 0$ at $r=0$ and~${\chi \to 0}$ as~${r \to
  +\infty}$. This means that the analogous particle starts at  
$r=0$ with zero velocity and some $\chi(0) = \chi_0$ and arrives
to~${\chi=0}$ as~${r\to +\infty}$. 

It is clear that $U_\omega(\chi)$ should be very special for the above
motion to occur, and this imposes conditions on the oscillon
amplitude $\chi_0$, frequency $\omega$, and the scalar
  potential~$V(\varphi)$. First,~${\chi=0}$ should be the maximum of  
$U_\omega$, or the particle would not stuck there 
as~${r \to +\infty}$. This simply
implies
\begin{equation}
  \label{eq:17}
  \omega < m\;,
\end{equation}
i.e.\ the binding energies~${\omega-m}$ of quanta inside the oscillon are
negative. Indeed, at small~$\varphi$ we can approximate $V(\varphi)
\approx m^2  \varphi^2 / 2$. This gives the canonical transformation
$\Phi$ of a harmonic oscillator and weak-field asymptotics of
the EFT form factors,
  \begin{equation}
    \label{eq:35}
     \Phi \approx \sqrt{2I / m} \, \cos\theta\,,  \quad
     h \approx m I\,,\quad 
    \mu_I \approx 4Im\,,\quad 
    U_\omega \approx m(\omega - m) \, \chi^2 \quad 
    \mbox{at small }I,\,\chi\;.
  \end{equation}
We see that the potential $U_\omega(\chi)$ has a maximum~${U_\omega
  = 0}$ at~${\chi=0}$ only if the frequency is bounded from the above,
$\omega < m$. 

Second, the analogous particle should start its motion 
downhill with positive mechanical energy (${d>1}$)  or with zero
energy (${d=1}$); otherwise it will not reach the maximum
$U_{\omega}(0)=0$. This gives necessary
requirements~${\partial_{\chi_0}U_\omega(\chi_0) > 0}$
and~${U_\omega(\chi_0) \geq 0}$, where $\chi_0 \equiv \chi(0)$.  In
terms of the original EFT fields,
\begin{equation}
  \label{eq:18}
  \omega > \Omega(I_0)\;, \qquad \qquad 
  \omega \geq h(I_0) / I_0\;,  \qquad \mbox{and} \qquad \mu_I (I)
  \ne 0 \;\; \mbox{at $I \leq I_0$}\;,
\end{equation}
where the oscillon amplitude $I_0 = \psi_0^2$ at $r=0$
  corresponds to $\chi_0$. Note that the ``EFT mass'' $\mu_{I}(I)$
  is non-negative by definition~(\ref{mu_coeff}), but
  may reach zero at singular points of the canonical transformation. Note
  also that in $d=1$ the mechanical energy is  
conserved and the second of Eqs.~(\ref{eq:18}) turns into an 
equality. We see that the oscillon frequency is bounded from the
below.

\begin{sloppy}

Together, Eqs.~\eqref{eq:17} and (\ref{eq:18}) give conditions for 
the oscillon with amplitude $I_0$ to exist in a given model:
\begin{equation}
  \label{eq:19}
   \Omega (I_0) < m\;, \qquad h(I_0)/I_0 < m\;, \quad\mbox{and}\quad
  \mu_I \big|_{I \leq I_0}   \ne 0  \qquad \mbox{for some $I_0$}\;.
\end{equation}
In the specific case $d=1$ we obtain an additional requirement
\begin{equation}
  \label{eq:21}
  \Omega (I_0) < h(I_0)/I_0 \qquad \qquad \mbox{in} \quad d=1\;.
\end{equation}
Now, recall that $\Omega(I)$ is the frequency of nonlinear
 oscillations in the potential~$V(\varphi)$, and $\Omega(0) = m$. The 
conditions~\eqref{eq:19}, mean that this frequency is 
{\it   smaller} at~${I = I_0}$ than at~${I\to  0}$, and the same
inequality holds for the amplitude-averaged frequency ${h(I_0) / I_0 =
I_0^{-1}\int_0^{I_0} \Omega(I)\, dI}$. We obtained a precise  way of
saying that the potential $V(\varphi)$ is attractive in the case of
strong fields. Typically, such a behavior is expected
from~$V(\varphi)$ which grows slower than $\varphi^2$ and has
decreasing $\Omega(I)$.  

\end{sloppy}

In $(1+1)$ dimensions, the friction term in Eq.~(\ref{eq:16}) is
absent and the oscillon profiles can be obtained
analytically. Let us illustrate this calculation in the
model~(\ref{U_def}) with~${m=1}$. We substitute the respective
form factors $h(I)$ and $\mu_I(I)$ [Eqs.~(\ref{eq:12}) and
(\ref{mu_coeff_th})] into the
conditions~(\ref{eq:18}), (\ref{eq:19}), and (\ref{eq:21}) and
  find out that the oscillons 
do exist in this model, their central amplitudes satisfy~${0 < I_0 <
  1}$,  and the frequencies belong to  the range~${1/2 < \omega <
  1}$. For these values, the ``mechanical'' potential $U_\omega(\chi)$
has the aforementioned specific form; it is plotted in
  Fig.~\ref{fig:mech_pot} at $\omega = 0.8$. Recall that the second
of Eqs.~(\ref{eq:18}) is an equality in one dimension. It fixes the
field in the oscillon center:~$\psi(0) = \sqrt{I_0} =
\sqrt{2-2\omega}$.

\begin{figure}[h]
  \centerline{\includegraphics{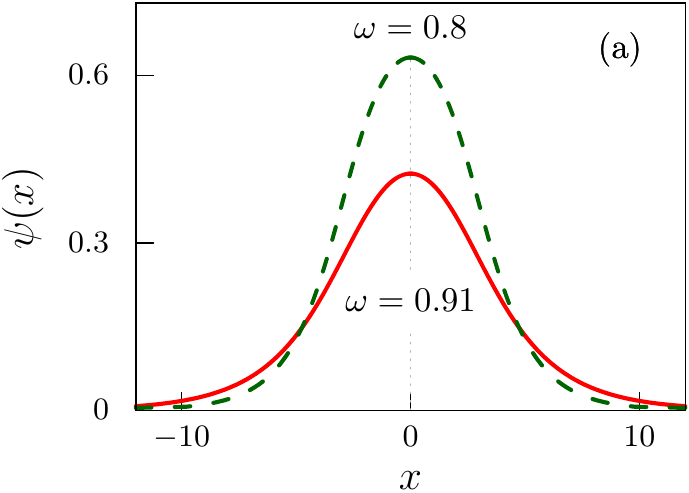}\hspace{5mm}
  \includegraphics{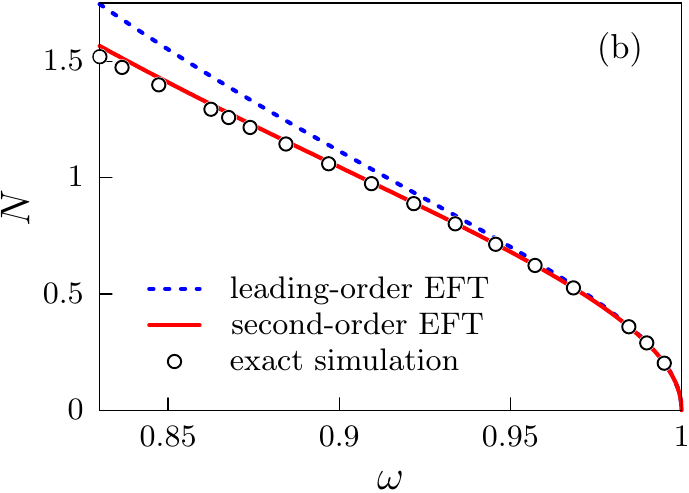}}
  \caption{(a) Two leading-order EFT profiles of one-dimensional
      oscillons in the model~(\ref{U_def}), see
      Eq.~(\ref{1d_analytic}). (b)~Oscillon charges $N(\omega)$  in 
      $d=1$ predicted by the leading-order EFT [dotted line,
        Eq.~(\ref{eq:30})], next-to-leading order EFT
      [solid line, Eq.~\eqref{eq:20}], and extracted from full
      numerical simulations (circles).  \label{fig:results-1d}}
\end{figure}

\begin{figure}
  \centering

  \vspace{1mm}
  \includegraphics{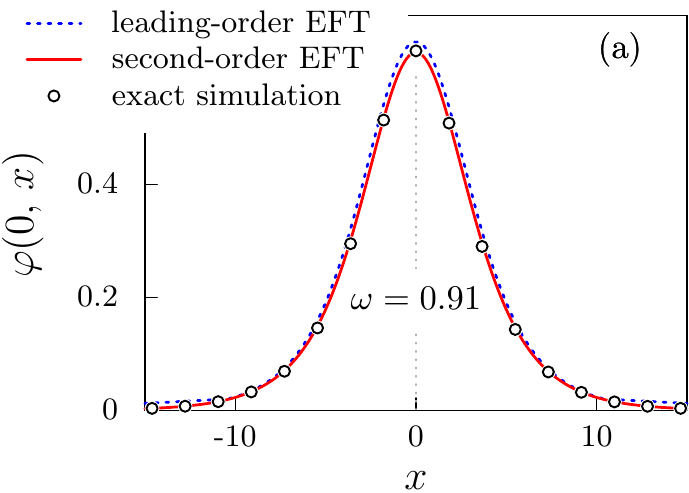}\hspace{5mm}
  \includegraphics{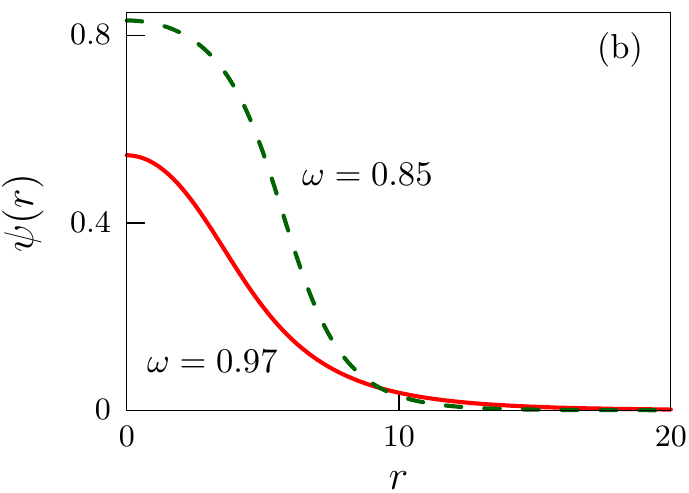}
  \caption{(a)~The field $\varphi(0,\, x)$ of one-dimensional oscillon
    with $\omega=0.91$ in the model (\ref{U_def}). Circles represent
    exact numerical simulations, dotted and solid lines 
    are the leading-order and second-order EFT results,
    respectively, see Eqs.~(\ref{tanh_action_angle}),
    (\ref{1d_analytic}), 
    and~(\ref{eq:44}). (b)~Two leading-order EFT profiles  $\psi(r)$ of
    three-dimensional oscillons in  the model (\ref{U_def}).}
  \label{fig:3d_profiles}
\end{figure}

Solving the profile equation, we obtain an analytic and even
function $\psi(x)$ in the implicit form,
\begin{equation}
  \label{1d_analytic}
  r = \frac{2}{\sqrt{2\omega-1}} \arctan
  \frac{\zeta(\psi)}{\sqrt{2\omega-1}} - \frac{1}{\sqrt{2\omega}}
  \arctan\frac{\zeta(\psi)}{\sqrt{2\omega}} + \frac{1
  }{\sqrt{2-2\omega}} \;
  \mathrm{arctanh}\,\frac{\zeta(\psi)}{\sqrt{2-2\omega}} \;,
\end{equation}
where $\zeta(\psi) = \sqrt{2 -2\omega - \psi^2}$ and $r = |x|$ in 
one dimension; see Appendix~\ref{sec:one-dimens-oscill} for
details. Together with~${\theta=\omega t}$, this solution specifies the
  oscillon field $\varphi(t,\, x) = \Phi(\psi^2,\, \theta)$ via
  Eq.~\eqref{tanh_action_angle}. We display $\psi(x)$ in 
  Fig.~\ref{fig:results-1d}a (solid and dashed lines) and related $\varphi(0, \, x)$ in 
  Fig.~\ref{fig:3d_profiles}a (dotted line). Note that the oscillon
  with higher~$\omega$ is lower in amplitude and has larger
  size~$R$. Indeed, Eq~\eqref{1d_analytic} shows that~$R \propto 
(1-\omega)^{-1/2}$ grows to  infinity\footnote{Also, $R$ is large
  at $\omega \to 1/2$. However, the respective
  profiles~(\ref{1d_analytic}) have forms of bubbles with
  almost constant interior fields $I \approx 1$ surrounded by thin
  walls with~${\partial_r I \sim  m I}$. Despite large size, such
  configurations break the EFT   conditions  (\ref{eq:6}),
  (\ref{eq:22}) and  therefore decay fast.} in the nonrelativistic
  limit $\omega \to   1$. In the latter case the EFT becomes exact.

Given $\psi(r)$, we calculate the charge $N$ and energy $E$ of
one-dimensional oscillons in the model~(\ref{U_def}) using
Eqs.~(\ref{eq:4}) and (\ref{F_generic}):
\begin{align}
  \label{eq:30}
  &N = \frac{4}{\sqrt{2\omega-1}} \; \arctan
  \frac{\sqrt{1-\omega}}{\sqrt{\omega-1/2}} -
  \frac{4}{\sqrt{2\omega}} \;
  \arctan \sqrt{1/\omega-1}\;,\\
  \label{eq:29}
  &E = \frac{4(1-\omega)}{\sqrt{2\omega-1}} \; \arctan
  \frac{\sqrt{1-\omega}}{\sqrt{\omega-1/2}} + 2\sqrt{2\omega} \;
  \arctan \sqrt{1/\omega-1}\;,
\end{align}
see the dotted lines in Figs.~\ref{fig:results-1d}b
and~\ref{fig:om_E}c,  and also Appendix~\ref{sec:one-dimens-oscill} for
details. Notably, $E(\omega)$ and $N(\omega)$ are monotonic in
$d=1$, which means that the 
  oscillons contain more charge at larger binding energies 
$m-\omega$. In Sec.~\ref{sec:Num_Tests}, 
we will see that this is no longer the case in three dimensions.

%%%%%%%%%%%%%%%%%%%%%%%%%%%%%%%%%%%%%%
\subsection{Longevity and stability}
\label{sec:long-stab-oscill}

\begin{sloppy}

Now, we make critical remarks on longevity and stability
of oscillons. One notices that the time derivative of the oscillon field
in Eq.~(\ref{eq:7}) equals ${\partial_t \varphi = \omega \,
  \partial_\theta\Phi}$ and does not coincide with the field
momentum  ${\pi_\varphi = \Pi \equiv \Omega \, \partial_\theta \Phi}$,
where the last identity is generically valid for mechanical action-angle
variables. This apparent mismatch is explained by observing
that~${\omega - \Omega(I)}$  should be small for the EFT to
work. Indeed, 
the second spatial derivative of $\psi(\boldsymbol{x})$ in
Eq.~(\ref{i_ball_profile_generic}) is proportional to $\omega - \Omega$ implying
that the oscillon size is proportional to $R \propto |\omega -
\Omega|^{-1/2}$. We conclude that $\Omega(I)$ should be almost
constant inside the oscillon,~i.e.
\begin{equation} 
  \label{eq:22}
  \mbox{EFT:} \qquad |d\Omega / dI| \ll \Omega / I  \qquad
  \mbox{or} \qquad \eta^2 \equiv |d\ln \Omega / d \ln I| \ll 1 \qquad 
  \mbox{at}
  \qquad I \leq I_0\;,
\end{equation}
where $I_0$ is the amplitude in the center. Moreover, one crudely
expects ${(mR)^{-2} \sim | d\ln \Omega / d\ln I|}$ because $\Omega -
\omega \propto \partial_I \Omega$,
cf.\ Eq.~(\ref{i_ball_profile_generic}). Note that we can relax   
Eq.~(\ref{eq:22}) a bit, e.g., by imposing it at $I_1 < I \leq I_0$,
where $I_1\ll I_0$. In this case the EFT works only in the oscillon
core~${I>I_1}$, and the entire object slowly evaporates through the 
boundaries. 

\end{sloppy}

Importantly, we expect the oscillons to live longer if Eq.~\eqref{eq:22}
is satisfied to a better precision. Indeed, charge conservation
prohibits their decay in the approximate EFT and, as we will see in
Sec.~\ref{sec:Corrections}, higher-order EFT  corrections do not ruin
this property. Hence, the oscillon lifetimes are large~--- presumably,
exponentially large~--- in the expansion parameter~$\eta\ll
  1$. This turns Eq.~(\ref{eq:22}) into a condition for the 
oscillon longevity. 

So far, we formulated all requirements in terms of mechanical
oscillation frequency~$\Omega$ in the scalar
potential~$V(\varphi)$. But once the appropriate $\Omega(I)$ is found,
one can restore~${\varphi \to -\varphi}$ symmetric potential using the
identity~\cite{LL1},
\begin{equation}
  \label{eq:23}
  \varphi(V) = \int_{0}^V \frac{dh}{\Omega(h) \sqrt{2V - 2h}}\;,
\end{equation}
where $\varphi(V)$ is the inverse to $V(\varphi)$ and $\Omega$ in the
right-hand side is expressed in terms of the  ``mechanical
energy''~$h(I)$. In particular, Eq.~(\ref{eq:22})
suggests that the scalar potential is nearly quadratic.

\begin{sloppy}

Let us discuss two obvious ways to satisfy all requirements. First,
the EFT always applies in the limit $I \to 0$ if $\partial_I
\Omega(0)$ is finite and nonzero. Indeed, any smooth
potential~$V(\varphi)$ is almost quadratic near the vacuum. The  respective
oscillons are long-lived and large in size: ${R\sim | m
  I_0 \partial_I \Omega(0)|^{-1/2} \propto   I_0^{-1/2}}$, where the
parameter in Eq.~(\ref{eq:22}) was exploited. Besides, due to
Eqs.~(\ref{eq:17}) and~\eqref{eq:18} their frequencies are 
slightly below the field mass, $m-\omega \propto I_0$. Finally, the
conditions \eqref{eq:19} for the existence of these objects fix the
sign~${\partial_I    \Omega(0) < 0}$. 

\end{sloppy}

A suitable technique working at  $I\to 0$ alongside the EFT is the
small-amplitude approximation~\cite{Dashen:1975hd,
  Kosevich1975, Fodor:2008es, Fodor:2019ftc}. In this case one
Taylor-expands all quantities in~$I$ and~$\varphi$. In particular, the
series for $\Omega(I)$ in  Eq.~(\ref{eq:23}) give,
\begin{equation}
  \label{eq:24}
  V = \frac12 m^2 \varphi^2   + \frac{g_3}{4} \,  \varphi^4 +
  \frac{g_5}{6} \, \varphi^6  + \dots\;,
\end{equation}
where $g_3 = \frac43 m^2 \partial_{I} \Omega(0) < 0$ and the dots are
terms with higher powers of $\varphi$. Apart from the necessarily
negative four-coupling $g_3$, this potential is generic~--- hence,
small-amplitude oscillons appear in all models with attractive
self-interactions. But notably, Eq.~\eqref{eq:24} is not applicable for
strong fields.  This is tolerable if Eq.~(\ref{eq:22}) is not satisfied
at large $I$ and if large-amplitude oscillons decay
faster. Our illustrative potential~(\ref{U_def}) with $\Omega = 1-I$
and~${m=1}$ is an example for that. However, there exist other cases
to consider. 

Second, in certain specifically chosen models $\Omega(I)$ may be flat
even for finite $I$. In this case the small-amplitude
expansion is not applicable. Let us take $\Omega = \Omega_0 + \delta
\Omega(I)$, where $\Omega_0$ is a constant and the function~${\delta 
\Omega(I) \ll  \Omega_0}$ is either bounded at large $I$ or grows
logarithmically.  Then the EFT
condition~(\ref{eq:22}) is satisfied. Expanding  
Eq.~(\ref{eq:23}) in~$\delta \Omega$, we arrive to 
\begin{equation} 
  \label{eq:31}
  V = \frac12\tilde{\Omega}^2(\varphi) \, \varphi^2\;, \qquad \mbox{with} \qquad
  \tilde{\Omega}^2  \approx \Omega_0^2 + \frac{2}{\varphi}
  \int\limits_{0}^{\Omega_0^2 \varphi^2/2} \frac{dh \,
    \delta\Omega(h)}{\sqrt{\Omega_0^2 \varphi^2 - 2h}}\;.
\end{equation}
This potential is almost quadratic since $\tilde{\Omega}(\varphi)$ is
nearly constant and grows at best logarithmically
at large $\varphi$. Notably, Eq.~(\ref{eq:31}) includes  monodromy
potentials in~\cite{Olle:2019kbo, Olle:2020qqy} which are already known to
support exceptionally long-lived large-amplitude oscillons. Our
estimates show that the size of the latter objects and their lifetimes can
be controlled by the parameter~(\ref{eq:22}),
cf.~Ref.~\cite{Olle:2019kbo}.

Next, we consider linear stability of oscillons. One generically
expects~\cite{Amin:2010jq} that these objects are destroyed by 
long-range perturbations if Vakhitov-Kolokolov
criterion~\cite{vk, Zakharov12}  is broken;
\begin{equation}
  \label{eq:25}
 \mbox{stability:} \quad dN(\omega) / d\omega = \omega^{-1} \;
  dE(\omega)/d\omega < 0\;.
\end{equation}
In the EFT, we rigorously prove this criterium in the simplest
possible way~\cite{vk, Zakharov12}, by demonstrating that the
oscillons can be the true 
minima of energy at a fixed  $N$ only if Eq.~(\ref{eq:25}) is
satisfied. We present the proof in
Appendix~\ref{sec:vakh-kolok-crit}. Note that 
one-dimensional oscillons in the model~(\ref{U_def}) do satisfy 
the stability criterion~(\ref{eq:25}), see Fig.~\ref{fig:results-1d}b.

One bewares~\cite{Amin:2010jq, Olle:2020qqy} that certain oscillons
may decay via parametric resonance~\cite{Tkachev:1987cd,
  Levkov:2020txo} for high-frequency 
modes that are discarded in the EFT. Certain intuition may
even suggest that this mechanism is efficient for large-size and
high-amplitude objects, i.e.\ precisely in the EFT case. Let us argue
that, to the contrary, parametric resonance 
is suppressed whenever the EFT condition  (\ref{eq:22}) holds. Indeed,
deep inside the oscillons high-frequency perturbations $\delta \varphi$ with wave
vectors~${k \sim O(m)}$ satisfy the equation [cf.~Eq.~\eqref{eq:5}], 
\begin{equation}
  \label{eq:32}
  (\partial_t^2  + k^2 ) \delta \varphi = -V''(\Phi(I_0,\, \omega t)) \,
  \delta \varphi\;,
\end{equation}
where the primes denote derivatives with respect to $\varphi$
and we ignored slow dependence on~$\boldsymbol{x}$ in
$I(\boldsymbol{x}) \approx I_0$. In the infinite  
medium, the periodic oscillations of~${V''}$ would always excite the modes inside some
resonance $k$-bands of width $\delta k \sim |V''_{\mathrm{osc}}|/ k$,
where Eq.~(\ref{eq:32}) was used in the estimate and the subscript
``osc'' denotes the oscillatory part of $V''$, cf.~\cite{LL1}. This makes $\delta
\varphi (t,\, \boldsymbol{k})$ grow exponentially within the bands. Instabilities in
compact objects, however, cannot develop unless the growing
modes can be localized inside them.  We obtain a crude condition for
the parametric 
instability~\cite{Tkachev:1987cd, Levkov:2020txo}:~${\delta k \gtrsim
  R^{-1}}$ or~${R |V''_\mathrm{osc}| \gtrsim m}$, where $R$ is the
oscillon size. However, the EFT condition (\ref{eq:22})  ensures that
the scalar potential is almost 
quadratic with suppressed oscillations of $V''$. We obtain ${(m
R)^{-2} \sim |d\ln \Omega / d \ln I| \propto |V''_{\mathrm{osc}}/m^2|}$   and
therefore small $R|V''_{osc}|/m \propto
(mR)^{-1} \ll 1$. Thus, parametric instability  is not expected to
occur if Eq.~(\ref{eq:22}) is satisfied, see also~\cite{Olle:2020qqy}.

%%%%%%%%%%%%%%%%%%%%%%%%%%%%%%%%%%%%%%%%%%%%%

\section{Application and numerical tests}
\label{sec:Num_Tests}
In two and three dimensions, the EFT profiles of oscillons $\psi(r)$
cannot be found analytically, so we compute them numerically in the
model~(\ref{U_def}). To this end we use the shooting
method\footnote{Namely, starting from
the initial data $\psi(0) = \psi_0$ and $\partial_r \psi(0) = 0$ at
$r=0$, we integrate Eq.~(\ref{i_ball_profile_generic}) with
coefficient functions~\eqref{eq:12},~\eqref{mu_coeff_th} to large $r$,
and then adjust the value of $\psi_0$ to select  localized solutions
with~${\psi \to 0}$ as~${r \to +\infty}$.}. The result
in $d=3$ is plotted in Fig.~\ref{fig:3d_profiles}b. As
expected, the profiles are lower and larger in size
if~$\omega$ is closer to $m=1$, and  the same property holds in two
dimensions. 

\begin{figure}
  \centering
  \includegraphics{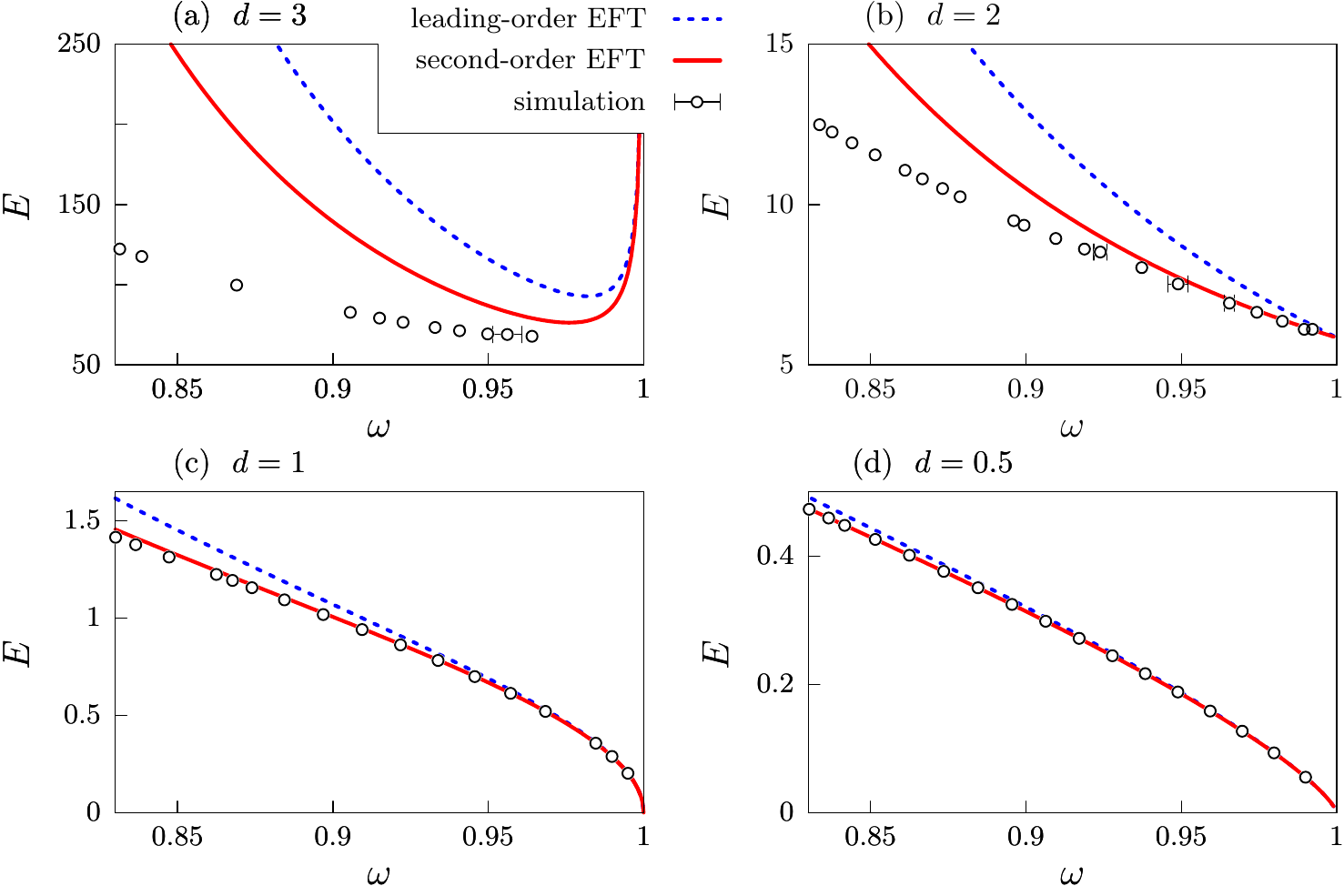}
  \caption{The energy $E(\omega)$ of $d$-dimensional oscillons
    in the leading-order EFT (dotted lines) and
    second-order EFT (solid lines). Circles with errorbars show results of
    exact numerical simulations.}
  \label{fig:om_E}
\end{figure}

The energies $E(\omega)$ of the EFT oscillons in various $d$ are
plotted with dotted lines in
Figs.~\ref{fig:om_E}a, b, and~c, see
Eqs.~(\ref{F_generic}) and (\ref{eq:4}). One observes a striking
feature:~$dE/d\omega$ is positive in the right-hand side of the
three-dimensional graph. This means that the criterion~(\ref{eq:25}) is
broken at~${\omega \approx m}$, $d=3$ and the respective oscillons are
unstable. Sadly, this happens  precisely in the region
where both the EFT and small-amplitude expansion work best.

\begin{sloppy}

In fact, oscillons with small amplitudes are generically unstable
in~${d\geq 3}$ dimensions. Indeed, in Sec.~\ref{sec:long-stab-oscill}
we estimated their sizes $R \propto 1/\sqrt{I_0}$ and
frequencies~${m-\omega  \propto I_0}$ in terms of the central amplitudes
$I_0$. This parametrically fixes their charges ${N(\omega) \propto
  R^d I_0 \propto (m-\omega)^{1-d/2}}$ and energies $E \approx m N$,
see Eqs.~(\ref{eq:4}) and~(\ref{E_N_Om_relations}). Thus, at~${\omega
\to m}$ the oscillon parameters $E$ and $N$ universally tend
to zero, remain constant, or grow to infinity in $d=1$, $2$, and
$d\geq 3$ dimensions, respectively. The stability criterion is broken
at $\omega \approx m$ in~${d\geq 3}$.

\end{sloppy}

Now, we compare predictions of the leading-order EFT with exact
numerical results in the model~(\ref{U_def}). We perform
spherically-symmetric 
simulations for $\varphi(t,\, r)$ like in
Sec.~\ref{sec:oscill-numer-ilustr}, but start from the approximate
EFT oscillons in Eq.~(\ref{tanh_action_angle}), where the profiles $I = \psi^2(r)$ are
computed above and $t = \theta = 0$. The latter configurations settle  
faster into the actual oscillons than the Gaussian initial
data. We wait at least $10^3$ oscillation cycles for that
to happen. Then we compute the periods~$T$ of the equilibrated objects as
the average time intervals between the consecutive maxima
of~$\varphi(t,\, 0)$. The oscillon frequencies are~${\omega = 2\pi /
  T}$, their energies are computed as  
\begin{equation} 
  \label{energy_osc}
  E = \int\limits d^d\boldsymbol{x} \,\left[ (\partial_t\varphi)^2 / 2
    + (\partial_i \varphi)^2/2 + V(\varphi) \right]\;,
\end{equation}
and the charges $N$ are given by Eq.~\eqref{eq:8}. Starting 
from various EFT profiles, we obtain oscillons with different
frequencies and functions $E(\omega)$, $N(\omega)$. 

In Appendix~\ref{Appendix:Num} we describe details of the above
numerical procedure, perform tests and estimate numerical errors
(which are small). One technical trick deserves to be mentioned in the
main text. In long simulations, we absorb all outgoing
radiation in  
order to prevent its reflection from the boundaries  onto the
central object. This is achieved by using\footnote{An
  alternative method is Kreiss-Oliger dissipation in~\cite{Honda:2001xg}.}
the ``artificial damping'' of  Ref.~\cite{Gleiser:1999tj}. Namely, we  modify
Eq.~(\ref{eq:5}) to
\begin{equation}
	\label{welcome_sponge}
	\left(\partial^2_t + H(r) \partial_t - \Delta \right) \varphi 
         = - V'(\varphi)\;, 
\end{equation}
where the ``sponge'' function $H(r)$ equals zero in the main part
$r< R_s$ of the lattice and smoothly increases at the lattice  
boundaries.  Equation~(\ref{welcome_sponge}) describes field in de
Sitter Universe with $r$-dependent  
Hubble constant  $H(r)/3$. In this setting, the outgoing scalar waves fade out
exponentially at~${r > R_s}$ like in the expanding Universe, while the
dynamics in the lattice center remains unmodified.  This strategy
allows us to control 
oscillons in long runs.

The simulation results for $N(\omega)$ in $d=1$ dimension
and\footnote{The functions
  $N(\omega)$ and $E(\omega)$ have similar shapes due to
  relation~(\ref{E_N_Om_relations}).} $E(\omega)$ in $d=1$, $2$,
and~$3$ are shown  in Figs.~\ref{fig:results-1d}b
and~\ref{fig:om_E}a,~b,~c  with circles. The errorbars represent 
numerical uncertainties whenever they exceed  the circle
size. One observes that the predictions of the leading-order EFT (dotted lines in the
figures) coincide with the exact results at $\omega \approx m=1$  but
deviate from them at 
lower~$\omega$, as can be expected from the criterion
(\ref{eq:22}). The only exception is the case~$d=3$ in  
Fig.~\ref{fig:om_E}a where the simulations cannot produce unstable
oscillons with~$\omega > 0.97$, and the agreement in 
the rest of the $\omega$ region is qualitative.

\begin{figure}
  \centering
  \unitlength=1mm
  \begin{picture}(70,50)
    \put(0,0){\includegraphics{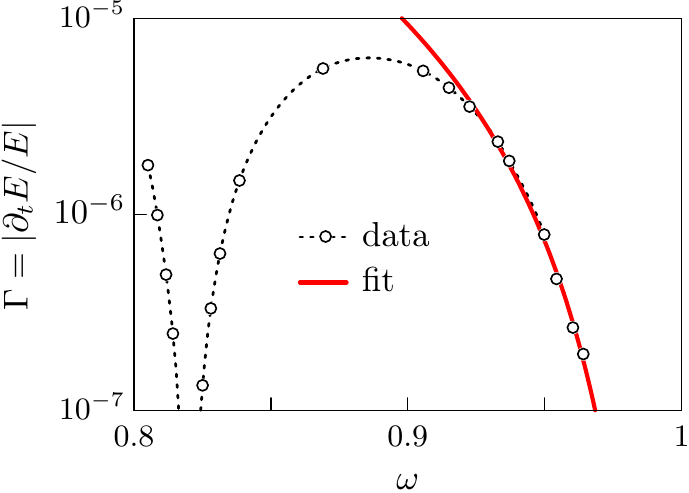}}
    \put(40,20.3){~(\ref{non_pert_order})}
  \end{picture}
    \caption{Decay rate $\Gamma(\omega) = E^{-1}\partial_t E$  of
      three-dimensional oscillons as a function of frequency in the
      model~(\ref{U_def}) (circles with an interpolating dashed
    line in the logarithmic scale). Solid line is the
      fit~\eqref{non_pert_order}.}
  \label{fig:gamma}
\end{figure}

In Sec.~\ref{sec:long-stab-oscill} we have made an important
suggestion that the oscillons live exponentially long
whenever the EFT applies. In other words, the emission rate
  $\Gamma$ of these objects is expected  to have the form
\begin{equation}
  \label{non_pert_order}
  \Gamma \equiv |E^{-1}\partial_t E| \sim \varsigma_1\,
  \mathrm{e}^{-\varsigma_2/\eta}\,,   
\end{equation}
where $\eta \equiv |d \ln \Omega / d \ln I |^{1/2} \sim (mR)^{-1}$ is
the EFT expansion parameter, $I = I(0)$ is the oscillon amplitude at
$r=0$, and $\varsigma_i$ are constants. Note that 
expression similar to Eq.~(\ref{non_pert_order})  was derived
analytically in the special case of small-amplitude
objects~\cite{Segur:1987mg, Fodor:2008du, Fodor:2009kf}. We confirm it
by computing the rates~$\Gamma(\omega)$ of three-dimensional oscillons
in full simulations, see the lines-points in Fig.~\ref{fig:gamma}
(logarithmic scale). It is clear that the data in the right part of
this graph are correctly described by the
exponent~\eqref{non_pert_order} (solid line), where the best-fit
parameters are~$\ln(\varsigma_1/m) = -5.55\pm 0.38$ and
${\varsigma_2 = 7.02 \pm 0.34}$ and $\eta =
\sqrt{I/(1-I)}$ in our model. We conclude that emission of oscillons
is indeed exponentially small at~${\eta \ll 1}$, and this corresponds
to the regime $\omega \approx 1$ in our model.

At smaller $\omega$ extremely long-lived oscillons with
  $\eta \sim O(1)$ appear. They are not controlled by the EFT and
deserve further study. Note that similar non-monotonic dependences of
emission rates were observed in~\cite{Zhang:2020bec, Olle:2020qqy,
  Cyncynates:2021rtf}.  

One sees another remarkable feature of Figs.~\ref{fig:om_E}a, b, and~c:
EFT works significantly better in lower dimensions. Indeed, it
reproduces exact results qualitatively in $d=3$ but becomes
almost precise for a range of frequencies in $d=1$. To test
this property, we notice that $d$ enters as a parameter into the
spherically-symmetric equations for $\varphi(t,\, r)$ and
$\psi(r)$. Hence, we can formally perform numerical 
calculations in non-integer $d$.  In Fig.~\ref{fig:om_E}d we
compare full 
simulations (circles) with the leading-order EFT (dotted line)
in~${d=0.5}$ and find almost perfect coincidence. This suggests that
the 
leading-order EFT becomes exact in the formal limit~$d\to 0$. In the
next Section we will find out why.

%%%%%%%%%%%%%%%%%%%%%%%%%%%%%%%%%%%%%%%%%%%%%

\section{The limit of zero dimensions}
\label{sec:limit-d-to-zero}
Now, we argue that oscillons turn into ever-lasting and
exactly periodic solutions in  the formal limit  $d\to  0$. This
explains why they are more common~\cite{Gleiser:2004an} in
lower dimensions.

We analytically continue the spherically-symmetric field equation 
\begin{equation}
  \label{field_eq}
  \partial^2_t \varphi - \partial^2_r \varphi - \frac{d-1}{r}\,
  \partial_r \varphi  = - V'(\varphi)
\end{equation}
to fractional $d$ imposing regularity condition $\partial_r \varphi =
0$ at the origin~${r=0}$. The latter requirement guarantees that  ${\varphi =
  \varphi_0(t) + \varphi_2(t)\, r^2+   O(r^4)}$ at small
$r$. Substituting the expansion into  Eq.~(\ref{field_eq})  and
setting $r=0$, we obtain equation for the first two coefficients:   
\begin{equation}
  \label{eq:39}
  \partial_t^2\varphi_0 - 2\varphi_2d = - V' (\varphi_0)\;.
\end{equation}
Notice that precisely in $d=0$  the field in the center
$\varphi_0(t)$ satisfies a mechanical equation which is independent
from the rest of the evolution. This agrees with the intuitive
interpretation of a  zero-dimensional field as a mechanical
pendulum~$\varphi_0(t)$ oscillating in the potential~$V(\varphi)$. In
our spherically-symmetric system the pendulum imposes time-periodic
Dirichlet boundary condition $\varphi(t,\, 0) = \varphi_0(t)$ on the
field in the bulk. Regardless  of the initial data,
the bulk configuration should eventually settle into a stationary
configuration oscillating with a period of the external force $\varphi_0(t)$. We
thus constructed an eternal exactly periodic solution in~$d=0$,
a prototype for the lower-dimensional oscillons. 

\begin{figure}
  \centering
  \includegraphics{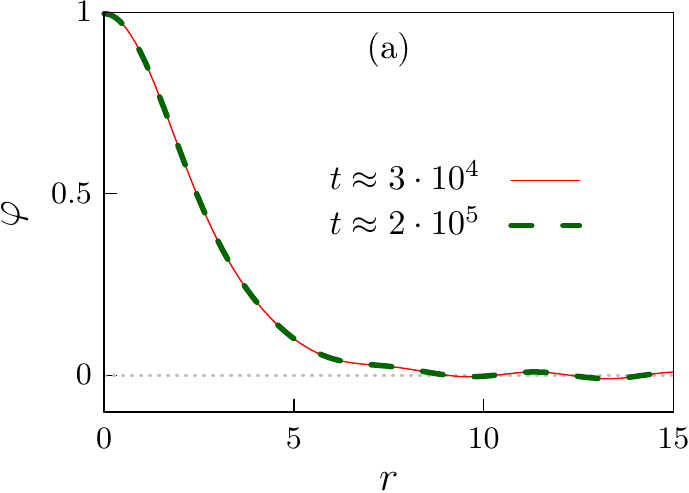}\hspace{5mm}
  \includegraphics{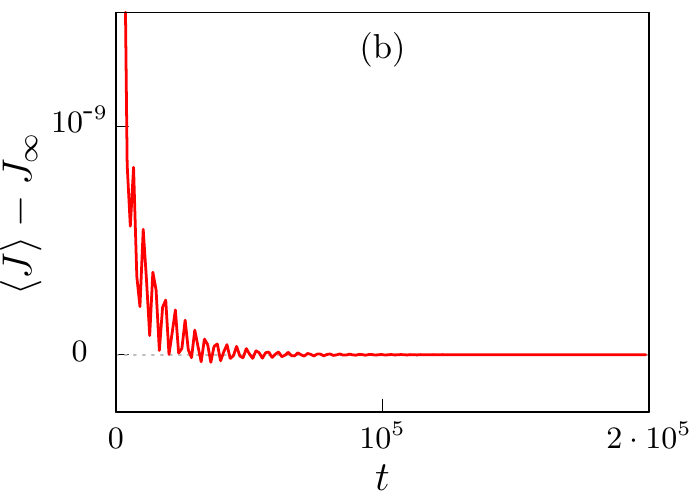}
  \caption{Numerical solution of Eq.~(\ref{field_eq}) with $d= 0$ in
    the model 
    (\ref{U_def}). The simulation starts from the approximate EFT
    oscillon with $\omega = 0.65$ and amplitude~$\varphi(0,0) \approx
    1$ in the center. (a)~Field configurations $\varphi(t,\, r)$ at two
    essentially different time moments  that both correspond to
    ${\partial_t  \varphi(t,\, 0) = 0}$. (b)~Energy flow $\langle
    J\rangle = - r^{-1}\langle \partial_t 
    \varphi \partial_r \varphi\rangle$ at~${r = 50}$ averaged
    over several oscillation periods; we subtracted the asymptotic value
    $\langle J\rangle \to J_\infty \approx 1.0088\cdot 10^{-5}$ as $t\to +\infty$.}
  \label{fig:zero_dimension}
\end{figure}

To see the periodic solution explicitly, we numerically
solve\footnote{Recall that in practice we solve
  Eq.~(\ref{welcome_sponge}) which includes an absorbing term at 
  large~$r$. In 
  $d=0$, this modification qualitatively affects the evolution by
  preventing growth of emitted radiation at~${r\to +\infty}$ and
  excluding the related nonlinear effects.}
Eq.~(\ref{field_eq}) with $d=0$ in the model~(\ref{U_def}). As before,
this evolution starts from the approximate EFT profile, say,
with~${\omega=0.65}$. After a stage of irregular wobbling, the 
solution enters the stationary regime with exact
period $2\pi/\omega$, and stays there to the very end of the
simulation at~${t\approx 2\cdot 10^5}$, see
Fig.~\ref{fig:zero_dimension}a.   

Consider now the limit $d\to 0$ of the integral quantities, e.g.\ the
energy (\ref{energy_osc}). We write the volume element as 
$d^d \boldsymbol{x}  = S_{d-1} (\mu r)^{d} \, dr/r$, where  ${S_{d-1}
  = 2\pi^{d/2} / \Gamma(d/2)}$ is the area of a unit sphere and the
scale $\mu$ is introduced to fix the mass dimension of 
the integrand. Notably,~${S_{d-1} \to 0}$ in the zero-dimensional
limit. In the total volume this multiplier is compensated by the
radial integral which diverges  logarithmically near $r=0$ in $d =
0$. As a consequence, all volume integrals in small $d$ 
receive main contributions near the origin. In particular, the energy  
(\ref{energy_osc}) takes the form,
\begin{equation*}
  E = E_0 + E_1d + O(d^2)\;,  \quad
  E_0 = (\partial_t \varphi_0)^2 / 2 + V(\varphi_0)\;, \quad
  E_1 = \int_{\epsilon}^\infty \rho \, dr/r + E_0 \ln (c\mu\epsilon)
\end{equation*}
where $\rho(r)$ is the integrand in Eq.~(\ref{energy_osc}), $c =
\mathrm{e}^{\gamma_E/2} \sqrt{\pi}  \approx 2.37$ is a number
involving the 
Euler constant~$\gamma_E$, and the limit $\epsilon \to +0$ is
assumed. We see, quite expectedly, that at small $d$ the leading
contribution comes from the pendulum energy $E_0$ in
the center, while the bulk part~$E_1d$ is suppressed. 

Nevertheless, the bulk energy is conserved even in $d=0$. In this
case Eq.~(\ref{field_eq}) gives~$\partial_t E_1 =  J_0 - J $, where
$J = -\lim\limits_{r\to \infty}r^{-1}\partial_t \varphi \partial_r
\varphi$ is the flux radiated to infinity and $J_0 =
-\partial_t \varphi \partial^2_r \varphi \big|_{r=0}$ comes 
from the origin. In Fig.~\ref{fig:zero_dimension}b we
plot the period-averaged flux~$\langle J\rangle$ for the
solution in Fig.~\ref{fig:zero_dimension}a. After several
oscillations, it approaches  a constant value $J_{\infty}$ 
indicating a stationary regime. We conclude that $\varphi_0$
oscillator sends $O(d)$ stationary energy flux  through the
bulk to infinity. 

It is already clear that our effective theory becomes exact in the
limit $d\to 0$. Indeed, we introduced $I$ and $\theta$ as the action and angle
variables in the mechanical system with potential~$V(\varphi)$. In
$d=0$, the EFT solution with $I(0) = I_0$ and $\theta = \omega t$ exactly
describes motion of the $\varphi_0$ pendulum at $r=0$. Moreover,  the
profile equation~(\ref{i_ball_profile_generic}) reduces at $r=0$ to
$\Omega(I_0) = \omega$ i.e.\ fixes the pendulum frequency to 
$\omega$. As a result, all corrections to the leading-order EFT come
from the bulk, and they are proportional to $d$. This is the reason
why the EFT is more precise in lower dimensions.

%%%%%%%%%%%%%%%%%%%%%%%%%%%%%%%%%%%%%%%%%%%%%

\section{Higher-order corrections}
\label{sec:Corrections}
In this Section, we evaluate EFT corrections and demonstrate that full
effective action has the form of a systematic gradient expansion, where 
every other term is suppressed as~$(mR)^{-2}$ with respect to the
previous one. To make the technicalities more transparent, we
  illustrate them in a toy model in Appendix~\ref{sec:Mech_Example}.

By itself, the transformation from $\pi_\varphi$, $\varphi$ to $I$,
$\theta$ leaves the theory exact. The main approximation of the 
leading-order EFT is the assumption that the latter
fields change slowly in space and time. In the second order, we split
them into smooth and  fast-oscillating parts:
\begin{equation}
	\label{split_smooth_fast}
	I = \bar{I} + \delta I\;, \qquad\qquad
        \theta = \bar{\theta} + \delta \theta\;, 
\end{equation}
where the perturbations have zero period averages,
$\langle \delta I \rangle = \langle \delta \theta \rangle = 0$, so
that~${\langle I\rangle = \bar{I}}$
and~${\langle\partial_t\theta\rangle = \partial_t
  \bar{\theta}}$. Below we will see that $\delta I$ and $\delta
\theta$ are small in the EFT expansion parameters. We will be able to
evaluate them perturbatively.

Now, the true EFT fields are $\bar{I}$, $\bar{\theta}$ and their
combination $\psi = \sqrt{\bar{I}} \cdot \mathrm{e}^{-i\bar{\theta}}$. An
effective action for them can be obtained by substituting the solutions
for  $\delta I$ and $\delta \theta$ into the exact action,   
\begin{equation}
  \label{aver_action}
  {\cal S} = \int dt d^d \boldsymbol{x} \left[ I\partial_t \theta - h(I) +
    \Phi \Delta \Phi/2\right]\;,
\end{equation}
where the canonical transformation $\varphi = \Phi(I,\, \theta)$ was
inserted into Eq.~(\ref{eq:9}). Varying
Eq.~(\ref{aver_action}) and subtracting the period-averaged equations,
we arrive to equations for perturbations,   
\begin{equation}
  \label{eq_corrections}
  \partial_t \delta I = j_\theta(I, \, \theta), \qquad \qquad 
  \partial_t \delta \theta = \Omega (I) - \langle \Omega \rangle -
  j_I(I,\, \theta)\;, 
\end{equation}
where, as usual, $\Omega = \partial_I h$, and we introduced  
the sources\footnote{Note that $j_\theta$ is $t\to -t$ antisymmetric
  and therefore has zero period average, $\langle j_\theta \rangle =
  0$.}
\begin{equation}
  \label{eq:41}
  j_\theta = \partial_\theta \Phi\, \Delta \Phi , \qquad \qquad 
  j_I = \partial_I \Phi\, \Delta \Phi - \langle \partial_I \Phi \, \Delta \Phi\rangle
\end{equation}
depending on $I = \bar{I} + \delta I$ and $\theta = \bar{\theta} +
\delta \theta$. We see that $\delta I$ and $\delta \theta$ are small,
indeed: in Eqs.~(\ref{eq_corrections}) they are determined by $j_I$ and
$j_\theta$ which include two spatial derivatives and hence are suppressed as
$(mR)^{-2}$.

In the second-order EFT, we solve for $\delta I$ and $\delta \theta$
in the leading order. In this case the time derivatives in
Eqs.~(\ref{eq_corrections}) can be changed to  ${\partial_t \approx
  (\partial_t\bar{\theta})\,  \partial_{\bar{\theta}}}$, since
$\bar{\theta}$ evolves progressively during the oscillation period.
Also, we can expand ${\Omega - \langle \Omega \rangle \approx \delta I
  \cdot\partial_{\bar{I}} \Omega}$ and  ignore the perturbations in $j_\theta$
and $j_I$. This gives, 
\begin{equation}
  \label{delta_I_theta_solut}
  \delta I \approx \frac{{\cal I}[j_\theta(\bar{I},\, \bar{\theta})]}{\partial_t
  \bar{\theta}}  \;, \qquad\qquad  
  \delta \theta  \approx \frac1{\partial_t {\bar{\theta}}}\,
  \Big\{\partial_{\bar{I}} \Omega(\bar{I}) \,
    {\cal I}[\delta I (\bar{I},\, \bar{\theta})]
    - {\cal I}[j_I(\bar{I},\, \bar{\theta})] \Big\}\;,
\end{equation}
where we introduced a natural primitive on the $\bar{\theta}$ 
circle for functions with zero average, 
\begin{equation}
  \label{I_primitive}
  {\cal I}[f] = \int_0^{\bar{\theta}} f(\bar{\theta}') \,d\bar{\theta}'  - \Big\langle
  \int_0^{\bar{\theta}} f(\bar{\theta}') \, d\bar{\theta}'
  \;\Big\rangle \qquad \mbox{satisfying} \quad \langle {\cal I}
                 [f]\rangle = \langle f\rangle = 0\;.
\end{equation}
From now on, the period averages $\langle \cdot \rangle$ are given by 
complete integrals over $\bar{\theta}$, like in Eq.~(\ref{eq:10}).

It is remarkable that Eqs.~(\ref{delta_I_theta_solut}) express
perturbations in terms of the EFT fields $\bar{I}(t,\,
\boldsymbol{x})$ and~$\bar{\theta}(t,\, \boldsymbol{x})$ 
which are arbitrary. Expanding Eq.~(\ref{aver_action}) in 
$\delta I$ and $\delta \theta$ and substituting the above solutions, we
arrive to the second-order effective action ${\cal S} \approx
{\cal S}_{\mathrm{eff}}^{(1)} + {\cal S}_{\mathrm{eff}}^{(2)}$,
where~${\cal S}_{\mathrm{eff}}^{(1)}[\bar{I},\, \bar{\theta}]$
is the leading-order result in Eq.~(\ref{action_eff_I}) and the correction
is 
\begin{equation}
  \label{F_corr_gen}
  {\cal S}_{\mathrm{eff}}^{(2)}[\bar{I},\, \bar{\theta}\,] = \int
  dt\,d^d \boldsymbol{x} \left[ \frac{1}{\partial_t \bar{\theta}}
    \left\langle j_I \mathcal{I}[j_\theta]\right\rangle -
    \frac{\partial_{\bar{I}} \Omega}{2 (\partial_t \bar{\theta})^2} \langle
    \left(\mathcal{I}[j_\theta]\right)^2\rangle\right]\;,
\end{equation}
see Appendix \ref{Append:Corrections_F:Gen} for the
detailed derivation. It is worth noting that $j_\theta$ and~$j_I$ in this
expression already depend on~$\bar{I}$ and~$\bar{\theta}$.

Let us make three important observations. First, the time integral in the
action effectively averages the Lagrangian over the oscillation
period. This can be done by integrating over
$\bar{\theta}$, so we added overall period averages to every term
of Eq.~(\ref{F_corr_gen}). As a consequence of the averaging, the
second-order EFT retains shift symmetry~$\bar{\theta} \to \bar{\theta} +
\alpha$, and a conserved charge  
\begin{equation}
  \label{eq:43}
  N = \int d^d \boldsymbol{x} \left[  \bar{I}  + \frac{\delta {\cal
        S}_{\mathrm{eff}}^{(2)}}{\delta (\partial_t \bar{\theta})}\right]\;,
\end{equation}
cf.\ Eq.~(\ref{eq:4}). This again guarantees that 
the stationary Ansatz $\bar{\theta} = \omega t$ and~$\bar{I} =
\bar{I}(\boldsymbol{x})$ passes the effective equations:
nontopological solitons~--- oscillons~--- exist in the next-to-leading order.

Second, the correction (\ref{F_corr_gen}) has $\partial_t
\bar{\theta}$ in the denominator~--- hence, EFT works for oscillons
with $\omega\sim m$, but not for static fields. Third, there
are four spatial derivatives of~$\bar{I}$ and~$\bar{\theta}$ in every
term of ${\cal S}_{\mathrm{eff}}^{(2)}$, they are hidden inside 
$j_\theta$ and~$j_I$. One can see this  explicitly: substitute
the sources~(\ref{eq:41}), expand the derivatives $\Delta \Phi(\bar{I},\,
\bar{\theta})$, move smooth quantities like~$\partial_i \bar{I}$ or
$\partial_i \bar{\theta}$ outside of the period averages, and
integrate the remaining coefficients in the effective Lagrangian
over~$\bar{\theta}$. In Appendix~\ref{Append:Corrections_F:Explicit}
we perform this  calculation in the simplified case of oscillon
with~$\partial_i \bar{\theta} = 0$ and $\bar{I} =
\psi^2(\boldsymbol{x})$. The result is
\begin{equation}
  \label{eq:46}
  {\cal S}^{(2)}_{\mathrm{eff}} = -\int  dt\,  F^{(2)}\,, \qquad
  F^{(2)} = - \int d^d {\boldsymbol{x}} \left[ d_1 \,
    (\partial_i \psi)^4 + d_2\, \psi
    \Delta \psi (\partial_i \psi)^2  + d_3 \,(\Delta \psi)^2 \right]\,,
\end{equation}
where $-F^{(2)}$ is a correction to the  Lagrangian and the
form factors $d_i$ depend on $\psi^2$. Say,
\begin{equation}
  \label{eq:40}
  d_3(\psi^2) = \left.\frac{4 \bar{I}}{\omega} \, \langle\,
  (\partial_{\bar{I}} \Phi)^2\, {\cal I} 
  [\partial_{\bar{\theta}} \Phi \partial_{\bar{I}} \Phi] \,\rangle
  - \frac{2 \bar{I}}{\omega^2} \,
  \partial_{\bar{I}} \Omega \, \langle \, \left({\cal 
    I} [\partial_{\bar{\theta}} \Phi \partial_{\bar{I}} \Phi ]
  \right)^2 \, \rangle\right|_{\bar{I} = \psi^2}\;,
\end{equation}
see Appendix~\ref{Append:Corrections_F:Gen} for other
coefficients. In any concrete model, the function $\Phi(\bar{I},\,
\bar{\theta})$ is known, so one can evaluate integrals over
$\bar{\theta}$. Analytic expressions for $d_{i}(\psi^2)$ in the
model~(\ref{U_def}) are given in
Appendix~\ref{Append:Corrections_F:Explicit}. 

As before, we determine the oscillon profile $\bar{I} \equiv \psi^2
(\boldsymbol{x})$, $\bar{\theta} = \omega t$ by extremizing the (minus) Lagrangian,
\begin{equation}
  \label{eq:42}
  F = F^{(1)} + F^{(2)} = E - \omega N\;,
\end{equation}
where the leading-order part $F^{(1)}[\bar{I}\,]$ and correction
$F^{(2)}[\bar{I}\,]$ are given by 
Eqs.~(\ref{F_generic}) and~(\ref{eq:46}), respectively. The last
identity in Eq.~(\ref{eq:42}) follows from the general
expression\footnote{Namely, $E   = F - \int d^d   \boldsymbol{x}  \,
  (\partial_t \bar{\theta}) \,\frac{\delta F}{\delta (\partial_t
      \bar{\theta})}$.} for energy and
Eq.~\eqref{eq:43}. Notably, the profile can be computed
perturbatively. One writes ${\bar{I} = \bar{I}^{(1)} + \bar{I}^{(2)}}$,
where 
 $\bar{I}^{(1)}$ extremizes $F^{(1)}$ and hence solves
Eq.~(\ref{i_ball_profile_generic}), and the correction~$\bar{I}^{(2)}$
satisfies linear equation 
\begin{equation}
  \label{I_1_small}
  \frac{\delta^2 F^{(1)}}{\delta \bar{I}^2} \cdot  \bar{I}^{(2)} = -
  \frac{\delta F^{(2)}}{\delta \bar{I}}
\end{equation}
in the background of $\bar{I}^{(1)}$. 
Once the profile is found, the oscillon field is given by Eq.~(\ref{eq:7}):
\begin{equation}
  \label{eq:44}
  \varphi(t,\, \boldsymbol{x}) = \Phi \left(\bar{I} + \delta I(\bar{I},\,
  \omega t),\, \omega t + \delta \theta (\bar{I},\, \omega t)\right)\;.
\end{equation}
Here $\delta I$ and $\delta \theta$ are provided by
Eqs.~(\ref{delta_I_theta_solut}) and $\bar{I}$ is the sum of 
$\bar{I}^{(1)}$ and $\bar{I}^{(2)}$. 

To illustrate calculation of the second-order oscillon field, we
numerically solved~${d=1}$ equations for $\bar{I}^{(1)}(x)$
and~$\bar{I}^{(2)}(x)$ in the model (\ref{U_def}) using the EFT
form factors in Appendix~\ref{Append:Corrections_F:Explicit}. The
second-order EFT field $\varphi(0,\, x)$ of one-dimensional oscillon with
$\omega = 0.91$ is plotted 
in  Fig.~\ref{fig:3d_profiles}a (solid line). As expected, it is
closer to the exact numerical result  (circles) than the
first-order EFT prediction (dotted line).  

It is worth noting that the mismatch between the derivative
$\partial_t \varphi$ and canonical momentum $\pi_\varphi \approx
\partial_t \varphi$ of the  oscillon is alleviated in the
second-order EFT. Indeed, taking the time derivative of
Eq.~(\ref{eq:44}) and using Eqs.~(\ref{delta_I_theta_solut}) and
(\ref{i_ball_profile_generic}), one obtains $\partial_t\varphi$
which is just~$O(mR)^{-4}$  different from the momentum $\pi_\varphi$
in Eq.~(\ref{eq:7}). Recall
that~$\partial_t \varphi - \pi_\varphi \sim O(mR)^{-2}$ in the
leading-order EFT~--- hence, these quantities coincide with increasing 
precision in higher orders.

On does not need the above corrections $\bar{I}^{(2)}$ and $\delta I$
for calculation of the second-order oscillon energy
$E(\omega)$  
and charge $N(\omega)$. Indeed, the
leading-order profile $\bar{I}^{(1)}$ extremizes~$F^{(1)}$ implying
\begin{equation}
  \label{eq:45}
  F [\,\bar{I}^{(1)} + \bar{I}^{(2)}\,] = F^{(1)} [\, \bar{I}^{(1)}\,]  +
  F^{(2)}[\, \bar{I}^{(1)}\,] + O(\,\bar{I}^{(2)} \,)^2 + O(\, \delta_{\bar{I}}
  F^{(2)}\, \bar{I}^{(2)}\, )\;,
\end{equation}
where both omitted terms are smaller than the next
correction to the effective action with six derivatives.  Once
$F(\omega)$ is obtained,  Eqs.~(\ref{E_N_Om_relations}) and~(\ref{eq:42})
give other oscillon parameters, 
\begin{equation}
  \label{eq:20}
  N =  - \partial_\omega {F}[\bar{I}^{(1)}]\;,
    \qquad \qquad
  E =F[\bar{I}^{(1)}] + \omega N\;, 
\end{equation}
which do not depend on $\bar{I}^{(2)}$ as well. In practice, the
$\omega$ derivative is computed numerically.

The charges $N(\omega)$ and energies $E(\omega)$ of the second-order
EFT oscillons in the model (\ref{U_def}) are shown with solid lines in
Figs.~\ref{fig:results-1d}b and~\ref{fig:om_E}. They are
closer to the results of exact simulations (circles) than the leading-order
predictions (dotted lines), and significantly so~--- in lower
dimensions and at $\omega \approx 1$. 

\begin{sloppy}

We finish this Section by remarking that the subsequent EFT corrections can
be calculated in the same way as the second-order one. In the third
order, one breaks the perturbations~$\delta I = \delta I^{(2)} +
\delta I^{(3)}$ and $\delta \theta = \delta \theta^{(2)} + \delta  
\theta^{(3)}$ into the leading parts ${\delta I^{(2)},\; \delta
\theta^{(2)} \sim O(mR)^{-2}}$ given by   
Eqs.~(\ref{delta_I_theta_solut}) and even smaller corrections~${\delta 
I^{(3)},\; \delta \theta^{(3)} \sim O(mR)^{-4}}$. The latter are
evaluated using Eqs.~(\ref{eq_corrections}) in the respective 
order; notably, the resulting expressions will 
include squares of the sources and hence four derivatives. Plugging
the refined perturbations into the exact action~(\ref{aver_action}),
one arrives at the next correction to the effective Lagrangian with six
derivatives. Importantly, the latter also should be averaged over
period, i.e.\ integrated over  $\bar{\theta}$. This means that
$\bar{\theta} \to \bar{\theta}+\alpha$  remains a symmetry at any EFT
order.

\end{sloppy}

%%%%%%%%%%%%%%%%%%%%%%%%%%%%%%%%%%%%%%%%%%%

\section{Comparison with small-amplitude expansion and automation}
\label{sec:limit-omega-to-zero}
We have already argued in Sec.~\ref{sec:oscill-effect-theory} that the sizes
$R$ of oscillons grow and the amplitudes $I_0$ become small when their
frequencies approach the field mass, $\omega \to m$. Namely, $R 
\sim O(\epsilon^{-1})$, $I_0 \sim  O(\epsilon^2)$, and $\varphi \sim
O(\epsilon)$ in terms of a small parameter $\epsilon^2 = m^2 -
  \omega^2$. In this case both EFT and small-amplitude
(nonrelativistic) expansion are applicable. The purpose of the present 
  Section is therefore twofold. On the one hand, we will demonstrate
  that all small-amplitude results  can be restored by re-expanding
  the effective action in the field strength to the respective order. On the
  other hand, we will see that the EFT can be realistically
  implemented in any model by symbolically computing long
  small-amplitude series for the form factors and summing them up numerically. 

With these purposes in mind, we consider a generic scalar
potential~\eqref{eq:24} in units with~$m=1$. The starting point of the
EFT is to introduce the mechanical action-angle variables in the potential
$V(\varphi)$. They can be evaluated 
automatically  in the form of half-integer series 
\begin{equation}
  \label{eq:phi_series_g}
  \Phi = \sqrt{2I}\,  \cos \theta + (2I)^{3/2} \, \frac{g_3}{32} 
  \left[ \, \cos (3 \theta) - 6 \cos\theta \,\right] + \dots\,,
   \quad h = I + \frac{3}{8} g_3 I^2  +  \gamma_3 I^3 + \dots\,,
\end{equation}
where the dots denote higher-order contributions, $g_i$ are the
coefficients of the potential~\eqref{eq:24}, and we defined $\gamma_3 
\equiv \frac{5}{12} \,g_5 - \frac{17}{64}\, g^2_3$. Recall 
that~$\Omega = \partial_I h$ and~$\Pi = \Omega \,\partial_\theta
\Phi$. Given the representation~(\ref{eq:phi_series_g}), we
calculate all~$\theta$ integrals in the EFT form factors. In
particular, 
\begin{equation}
  \label{eq:47}
  \frac{1}{\mu_I} = \frac{1}{4I} - \frac{9}{16}\, g_3  + O(I) \;,
  \qquad \qquad d_3 = \frac{1}{8\omega} + O(I)\;,
\end{equation}
where Eqs.~\eqref{mu_coeff} and \eqref{eq:40} were exploited. Unlike in
the previous Section, now we use $I$ and $\theta$ for smooth EFT
fields, omitting the overbar.  It is clear that arbitrarily long series for the
EFT form factors can be symbolically computed using the series for $\Phi$. 

\begin{sloppy}

Next, we use the EFT to obtain small-amplitude expansion for the
oscillon with ${I = \psi^2 (\boldsymbol{x})}$, ${\theta = \omega
  t}$, and $\omega \approx 1$. We substitute the
series~(\ref{eq:47}) into the 
second-order effective  Lagrangian~\eqref{F_generic}, \eqref{eq:46}
and reshuffle its terms according to the power-counting in~${\epsilon \ll
  1}$, 
\begin{multline}
  \label{eq:49}
  F^{(1)} + F^{(2)} = \int d^d \boldsymbol{x} \left\{ 
  \frac12\,  (\partial_i\psi)^2 + \frac{\epsilon^2}{2} \, \psi^2   + \frac{3}{8} \,
    g_3 \psi^4 \right.\\
    \left. - \frac{ 9}8 \,  g_3 \psi^2 (\partial_i \psi)^2
    + \frac{\epsilon^4}{8} \, \psi^2
    +  \gamma_3 \psi^6  - \frac{1}{8} \, (\Delta
    \psi)^2 + O(\epsilon^8)\right\}\;,
\end{multline}
where the first and second lines are $O(\epsilon^4)$ and 
$O(\epsilon^6)$, respectively, since $\psi = O(\epsilon)$,
${\partial_i \psi \sim O(\epsilon^2)}$, and $1-\omega \approx
\epsilon^2/2 + \epsilon^4/8$.  As before, the oscillon
profile can be found by varying Eq.~(\ref{eq:49}) with respect to 
$\psi(\boldsymbol{x})$. This time, we evaluate it perturbatively in  
$\epsilon$, i.e.\ search for solution in the form, 
\begin{equation}
  \label{eq:52}
  \psi   =  \frac{\epsilon}{\sqrt{2}}\, p_1 (\boldsymbol{x}) +
  \frac{\epsilon^3}{4\sqrt{2}} 
  \left[4p_3(\boldsymbol{x}) - p_1 (\boldsymbol{x})\right] + O(\epsilon^5)\;, 
\end{equation}
where $p_1$ and $p_3$ control the leading-order and second-order
contributions, respectively. In the first two orders, we obtain equations
\begin{equation}
  \label{eq:54}
  \epsilon^{-2}\Delta p_1 -p_1 - \frac{3}{4} g_3 p^3_1 = 0\;, \qquad
  \epsilon^{-2}\Delta p_3 - p_3 - \frac{9}{4} \, g_3  p^2_1  p_3 =
  \left(\frac{5g_5}{8} + \frac{3g_3^2}{128}\right) p^5_1\;.
\end{equation}
Notably, Eqs.~(\ref{eq:54}) literally coincide with equations appearing in the
small-amplitude expansion in~\cite{Fodor:2008es}. The oscillon
field can be then found by expanding Eq.~(\ref{eq:44}) in $\epsilon$,
\begin{equation}
  \label{eq:55}
  \varphi(t,\, \boldsymbol{x}) = (\epsilon p_1 + \epsilon^3  p_3)
  \cos(\omega t) + \frac{g_3}{32} (\epsilon p_1)^3\cos(3\omega t) +
  O(\epsilon^5)\;,
\end{equation}
again in agreement with~\cite{Fodor:2008es}. Finally,
the power-law expansions for $E(\omega)$ and $N(\omega)$ in~${\epsilon^2 =
1-\omega^2}$ follow from Eqs.~(\ref{eq:49}), (\ref{eq:52}), 
and~\eqref{eq:43}. 

\end{sloppy}

\begin{figure}[h]
\centering
\includegraphics{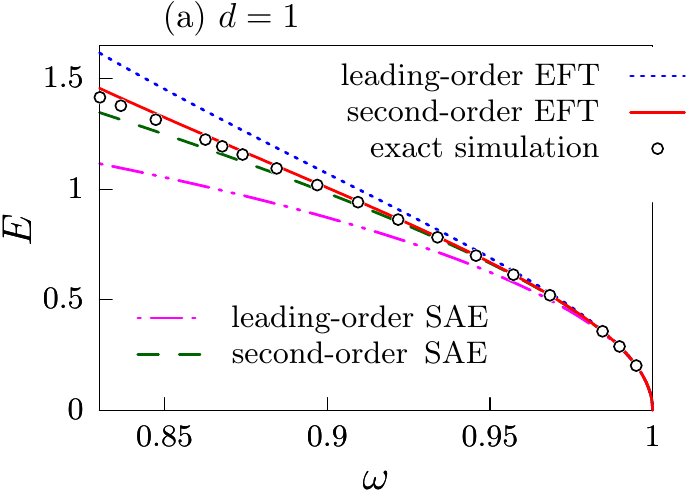}\hspace{5mm}
\includegraphics{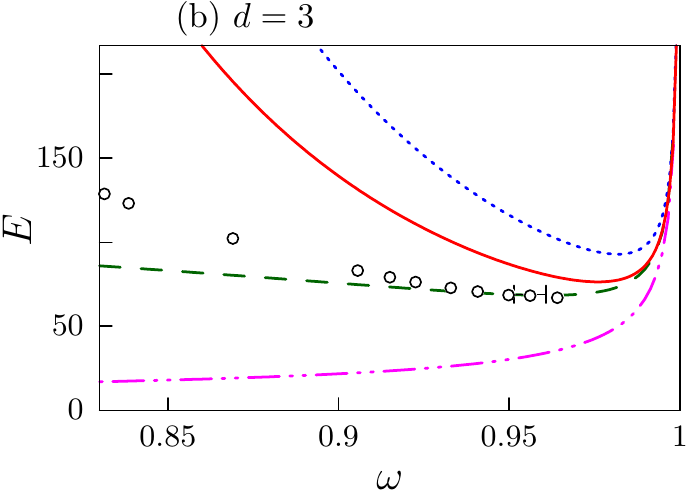}
\caption{Oscillon energies $E(\omega)$ in the
  model~(\ref{U_def}) as predicted by two orders of the EFT
  (dotted and solid lines) and two orders of the small-amplitude
  expansion (SAE, dash-dotted and dashed lines). Theoretical
  predictions are compared with the results of exact numerical
  simulations (circles with errorbars). Figures (a) and (b)
  correspond to $d=1$ and $3$ dimensions, respectively.}
\label{fig:sae-compare}
\end{figure}

It is worth reminding that the EFT is not equivalent to the small-amplitude
approach even at~${\omega \approx 1}$. Indeed, these methods
have different expansion parameters:~$R^{-1} \to 0$ at
finite~$\varphi$ (EFT), and $R^{-1} \sim \varphi \to 0$ 
(small-amplitude). To compare the approximations, we solved
Eqs.~(\ref{eq:54}) in the 
model~(\ref{U_def}) and obtained the oscillon energies
$E(\omega)$ in the first two orders of 
small-amplitude expansion.  In Figs.~\ref{fig:sae-compare}a
and~\ref{fig:sae-compare}b we 
compare these results (dash-dotted and dashed lines, respectively) with the
leading-order and second-order EFT predictions  (dotted and
solid lines), as well as with the results of
exact numerical simulations (circles with errorbars). We consider the
cases of $d=1$ and~$3$ dimensions. It is clear that the EFT works better in
lower $d$ but becomes less precise than small-amplitude expansion
in $d=3$. On the other hand, even in the leading order it predicts
two~--- stable and unstable~--- branches of three-dimensional
oscillons [cf.\ Eq.~(\ref{eq:25})], which is qualitatively
correct. The small-amplitude expansion incorrectly suggests instability of
all three-dimensional oscillons in the leading order, but becomes more
precise in higher orders. We believe that these
properties remain valid in other generic models.

%%%%%%%%%%%%%%%%%%%%%%%%%%%%%%%%%%%%%%%%%%%%%
\section{Discussion}
\label{sec:conclusions}
In this paper, we have developed a consistent Effective Field Theory (EFT)
description of scalar oscillons with arbitrary amplitudes
in the limit of large size. The main trick of the method is to perform
a canonical  transformation from the original field $\varphi$ and its
momentum~$\pi_\varphi = \partial_t \varphi$ to long-range
variables~$\psi(t,\, \boldsymbol{x})$  and~$\psi^*(t,\,
\boldsymbol{x})$. Once 
this is done, the effective classical action for smooth parts of $\psi$ and $\psi^*$
takes the form of a systematic gradient expansion with more spatial
derivatives appearing in every next term. The sole parameter of the
expansion is~$(mR)^{-2}$, where $m$ is the field mass and $R$ is the
spatial scale. At any order, such an effective theory has a global~U(1)
symmetry $\psi \to \psi\,  \mathrm{e}^{-i\alpha}$ and, as a
consequence, a family of nontopological solitons called oscillons. In
the original, exact model, the $U(1)$ symmetry is broken by 
the nonperturbative to EFT effects~--- hence, the lifetimes of large-size
oscillons are expected   to be exponentially long in~$(mR)^2$. 

It is remarkable that all presently known long-lived oscillons have
sizes $R$ exceeding the inverse field mass $m^{-1}$ by 
factors of few  or even parametrically~\cite{Fodor:2008es, Amin:2010jq,
  Olle:2019kbo, Olle:2020qqy}~--- at least, we are not aware of 
notable  exceptions from this rule. Thus, conservation of the~U(1)
charge~\cite{Kasuya:2002zs, Kawasaki:2015vga} in the EFT may be
universally responsible for the oscillon longevity. Importantly, 
the effective theory comes with conditions on the scalar
potential~$V(\varphi)$ needed for the existence of long-lived
oscillons. We formulated them in terms of a fictitious
mechanical motion  $\varphi = \varphi (t)$ in the potential $V(\varphi)$ or, more
specifically, in terms of the oscillation frequency
$\Omega(I)$ as a function of  the amplitude $I$~--- the
action variable in this mechanical system. 
Stationary solitons (oscillons)  exist in the EFT only if~$\Omega(I_0) < m$  
for some~$I_0$, see Eqs.~\eqref{eq:19}, and
\eqref{eq:21} for the complete list of conditions. Besides, the oscillons
are large and hence long-living if the mechanical frequency is
almost constant:~$|\partial_{I}\Omega| \ll \Omega / I$ at~$I \leq I_0$,
cf.\ Eq.~\eqref{eq:22}. In simple terms, these conditions mean that
the potential~$V(\varphi)$ is  attractive and nearly quadratic, see 
also~\cite{Olle:2020qqy, Cyncynates:2021rtf}. Finally, 
the oscillons are linearly stable\footnote{Note also that parametric decay
  of oscillons into high-frequency modes is inefficient once the EFT conditions are met, see
  Sec.~\ref{sec:long-stab-oscill} and cf.~\cite{Olle:2020qqy}.} with
respect to long-range  perturbations if the Vakhitov-Kolokolov 
criterion~\eqref{eq:25} is fulfilled.

We discussed two obvious ways to satisfy all requirements.  First, 
the conditions are generically met in  the small-amplitude limit  $I\to 0$ if
$V(\varphi)$ is attractive\footnote{Note that small-amplitude oscillons are linearly
  stable only in $d\leq 2$ dimensions, 
  see Sec.~\ref{sec:Num_Tests}.}, see
Eq.~\eqref{eq:24}. This is the case 
when the standard nonrelativistic (small-amplitude)
expansion~\cite{Dashen:1975hd, Kosevich1975, Fodor:2008es,
  Fodor:2019ftc} is applicable on par with the EFT. Second, the
conditions are fulfilled if ${V = \tilde{\Omega}^2 \,\varphi^2/2}$ and the 
function~$\tilde{\Omega}(\varphi)$ is almost constant in the sense of 
Sec.~\ref{sec:long-stab-oscill}. The latter option includes, in particular,
monodromy potentials $V\propto \varphi^n$ with $n\approx 2$ which are
known to support large-amplitude and exceptionally long-lived 
oscillons~\cite{Olle:2019kbo, Olle:2020qqy}. It would be
interesting to investigate such objects within the EFT.
 
The above two options do not exhaust all possibilities. In general,
one may construct the appropriate function~$\Omega(I)$ 
and then restore the potential
$V(\varphi)$ using Eq.~\eqref{eq:23} (in the symmetric
case). The resulting model should support long-lived oscillons. 
In particular,~$\Omega (I)$ can be flatter in some range of~$I$'s or have a
minimum at some value. Then the evaporation rates of the
respective oscillons should be non-monotonic in 
amplitude and frequency. Similar non-monotonic effects were numerically
observed in~\cite{Zhang:2020bec, Olle:2020qqy,
  Cyncynates:2021rtf}, and some of them may
correspond to features of $\Omega(I)$. 

One may find calculation  of the effective action  technically compelling
in models with nontrivial potentials.  But in fact, this procedure can
be automated using computer algebra. Namely, the 
transformation to $\psi$ and~$\psi^*$ and all coefficients in the
effective action
can be computed symbolically in the form of long
power-law series, which can be then summed up numerically. We discuss
this possibility in Sec.~\ref{sec:limit-omega-to-zero}. Using the same
trick, one may try to apply the EFT to models with nontrivial field content,
e.g.~\cite{Zhang:2021xxa}.

We performed extensive tests of the
EFT. First, we demonstrated that the standard small-amplitude series
for oscillons~\cite{Fodor:2008es} can be exactly reproduced by
expanding the effective action in fields, see
Sec.~\ref{sec:limit-omega-to-zero}. Second, we compared the EFT
predictions with the exact numerical simulations in the model with a
plateau potential. Quite expectedly, the effective technique becomes
precise in the case of large-size and  long-lived objects. Besides, it
works better in lower dimensions.

We believe that the classical EFT may have wider region of applicability
than just field theoretical oscillons. In particular, it may be
helpful for studying dynamical systems, see the toy example
in Appendix~\ref{sec:Mech_Example}. 

A fascinating project for the future is analytical calculation of
the oscillon evaporation rates in the limit  $R \to \infty$ for
arbitrary field strengths. In Sec.~\ref{sec:Num_Tests} we
  numerically confirmed that these rates are exponentially small in
  the EFT expansion parameter~$\eta \sim (mR)^{-1} \ll 1$. However, 
  analytic derivation of the oscillon lifetimes is far beyond the scope of
  the present paper. Presumably, this can be done by  extracting
exponentially small,  complementary to the EFT parts of the oscillon
solutions via complex analysis, like in the  case of small
amplitudes~\cite{Segur:1987mg, Fodor:2008du,Fodor:2009kf,
  Fodor:2019ftc}.

Yet another direction for future research is related to  formal limit of
zero spatial dimensions,~$d\to 0$. Recall that spherically-symmetric field
equation for $\varphi(t,\, r)$ remains nontrivial in $d=0$. In this
paper we demonstrated that the respective ``zero-dimensional''
oscillons are 
exactly periodic and eternally living. This explains, why oscillating
objects appear  more frequently and live longer in lower
dimensions~\cite{Gleiser:2004an}. One may try to approximate 
the lower-dimensional objects with zero-dimensional ones and then
compute corrections in $d$.

Notably, our leading-order effective
action  becomes exact in 
$d= 0$, since all higher-order terms in the derivative expansion are
proportional to $d$. Accordingly, the EFT works better in
lower dimensions. 

\acknowledgments
This work was supported by the grant
RSF 22-12-00215. Numerical calculations were performed on the
Computational Cluster of the Theoretical Division of~INR RAS.

%%%%%%%%%%%%%%%%%%%%%%%%%%%%%%%%%%%%%%%%% 
\appendix

%%%%%%%%%%%%%%%%%%%%%%%%%%%%%%%%%%%%%%%%%%%%%%%

\section{Numerical methods}
\label{Appendix:Num}

\begin{sloppy}

We solve the spherically-symmetric equation for $\varphi(t,\, r)$
on a uniform radial lattice with~${N_r+1}$ sites ${r_j =   j \Delta
  r}$ in a finite box~${0 \leq r \leq L_r}$, ${\Delta r = 
  L_r/N_r}$. The field values~${\varphi_j = \varphi(t,\,
r_j)}$ are stored on the lattice sites. In practice, $L_r = 750$ and~$375$
in $d\geq 1$ and~${d<1}$ dimensions, respectively. Besides,  we use 
${N_r = 16384}$ sites in ${d<3}$, $N_r = 32768$ 
in $d=3$, and repeat computations at twice  smaller~$N_r$ to
control the discretization errors.

In the Hamiltonian form and in spherical symmetry, the field
equation~(\ref{eq:5}) reads,
\begin{equation}
  \label{eq:first_order}
  \partial_t \varphi =  \pi_\varphi\;, \qquad
  \partial_t \pi_\varphi = R[\varphi]\;, \qquad \mbox{where} \quad
  R \equiv \partial^2_r \varphi + \frac{d-1}{r} \, \partial_r\varphi -
  V'(\varphi)\;.
\end{equation}
We impose Neumann boundary conditions $\partial_r \varphi = 0$ 
at the lattice
boundaries $r=0$ and~$L_r$.  Discretization of spatial derivatives in
Eq.~(\ref{eq:first_order}) is done using fast Fourier
transform (FFT). 
Namely, we Fourier-transform the field, ${\varphi(r) =
  \tilde{\varphi}_0 + \sum_k \tilde{\varphi}_k \cos(p_k  r)}$, where
$\tilde{\varphi}_k$ is the image at discrete momenta~${p_k = \pi 
k/L_r}$, and compute the derivatives by acting with $\partial_r$
and~$\partial_r^2$ on this representation. Say, the first
derivative is given by the sum ${\partial_r \varphi(r_j)  = -\sum_k p_k
\tilde{\varphi}_k\sin(p_k r_j)}$ which is easily computed. In this way
we get the right-hand side $R(r_j) = R^{(j)}$ in Eqs.~(\ref{eq:first_order})
with exponentially small accuracy ${\delta R^{(j)} \sim
\exp(-\mbox{const}/\Delta r)}$: all numerical errors in the FFT
procedure come from large momentum cutoff ${p_k   <p_{\max} \equiv \pi
  /\Delta r}$ cropping exponentially small tails of Fourier images. In
practice we exploit the FFTW3 library~\cite{Frigo:2005zln}.

\end{sloppy}

We evolve Eq.~(\ref{eq:first_order}) in time  using fourth-order
symplectic  Runge-Kutta-Nystr\"om (RKN4) integrator~\cite{McLachlan:PDE,
  Regan:Neuman}. In this method every time step consists of four
sequential replacements $\pi_\varphi \to \pi_\varphi +  a_\lambda \, 
\Delta t \, R[\varphi]$, $\varphi \to \varphi + b_\lambda \, \Delta 
t \,\pi_\varphi$, where ~$\lambda = 1\dots 4$ and the parameters
$a_\lambda$ and $b_{\lambda}$ are given in~\cite{McLachlan:PDE,
  Regan:Neuman}; in particular, $\sum a_\lambda = \sum b_\lambda = 1$. 
Overall, this gives $O(\Delta t^4)$ precision and exact conservation
of a symplectic form. In calculations, we set~${\Delta t =10^{-2}}$ and
use twice larger time step for numerical tests. All our parameters
satisfy the stability criterion~\cite{Regan:Neuman}~${\Delta t \lesssim
0.97 \, \Delta r}$ of the RKN4 method.

Now, recall that we modify the field equation with the sponge  $H(r)$
to absorb the outgoing radiation, see Eq.~\eqref{welcome_sponge}. The new 
term is incorporated into the RKN4 procedure by changing
the~$\pi_{\varphi}$ replacements to $\pi_\varphi \to (1+\frac12
a_\lambda H \Delta t)^{-1} \left\{ (1-\frac12 a_\lambda H
\Delta t)\, \pi_\varphi + a_\lambda \Delta t R[\varphi] \right\}$. We
set~${H = 0}$ at $r < R_s$ and $H = H_0 \cdot (r - R_s)^2$,
$H_0 = 10^{-6}$ outside of this sphere. In~${d\geq 1}$ dimensions,
we cover roughly one half of the
lattice with the absorbing region:~$R_s = 400$. In~${d<1}$ the emitted
linear waves {\it    grow} 
at large $r$, so we move the sponge closer,~${R_s = 50}$.   

Finally, we obtain the energy $E$ and charge $N$ of oscillons
discretizing the 
integrals~\eqref{energy_osc}, \eqref{eq:8} in the second 
order. It is worth noting that in fractional~$d$ the volume factors
in the integrands have soft
singularities $r^{d-1}$   which are explicitly accounted 
for. Also, we restrict all  integrations to the regions $r < 
r_{\max}$ dominated by oscillons. Here~$r_{\max}$ is the radius
at which the energy and charge densities stop falling off
exponentially. 

\begin{figure}
  \centering
  \includegraphics{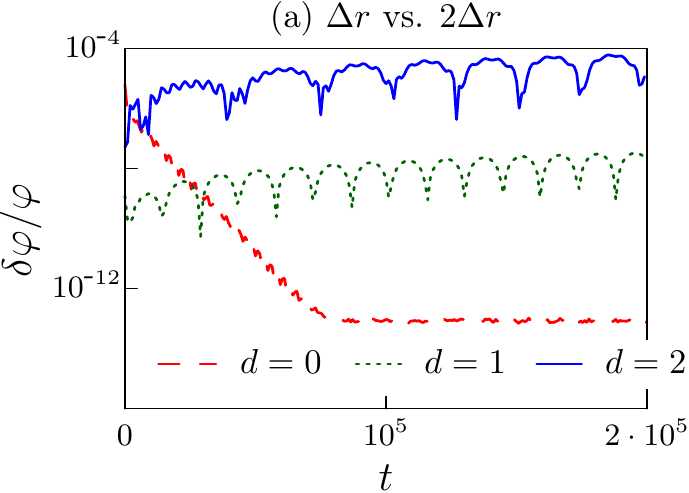}\hspace{5mm}
  \includegraphics{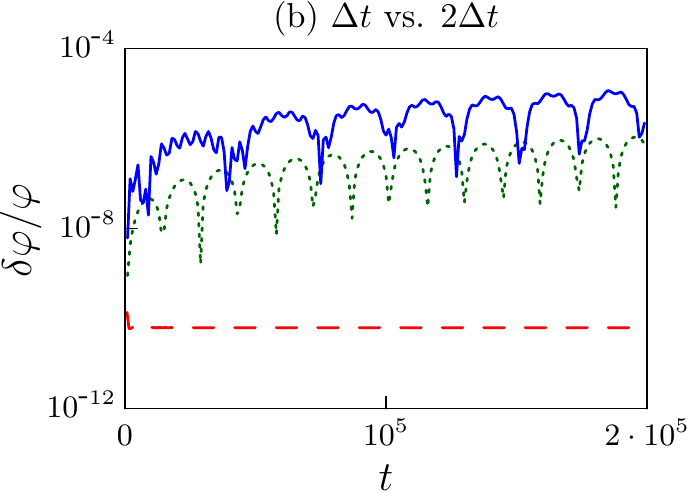}
  \caption{Maximal relative difference $\delta \varphi / \varphi
    \equiv \max_{r}|\delta \varphi| / \max_{r} |\varphi|$ between the
    two numerical
    solutions with~${\omega=0.99}$ which are computed (a)~on grids with $N_r =
    16384$ and $8192$ i.e.\ at twice different $\Delta r$; (b)~with
    time steps $\Delta t = 10^{-2}$ and $2\cdot 10^{-2}$.}
  \label{fig:diff}
\end{figure}

The above algorithm remains stable even during long simulation runs
up to ${t = 2\cdot 10^{5}}$. We estimate the numerical errors by
changing the lattice spacing~$\Delta r$ and time step  
$\Delta  t$. In this way, we checked that the discretization errors are
exponentially sensitive to~$\Delta r$ and proportional to~$(\Delta
t)^4$. The sizes of these two inaccuracies are roughly
comparable. Typically, they are orders of magnitude below  
the relative level of $10^{-4}$ and reach this value in the worst\footnote{Recall that we
  switch to twice larger $N_r$ and better accuracy in~${d=3}$
  dimensions.} case~${d=2}$, see 
Fig.~\ref{fig:diff}. Finally, we checked energy
conservation\footnote{More precisely, we test the law $\partial_t E =
  -J$, where $E$ is the energy in the box $0\leq r \leq R_s$ and the
  out-flux $J = -S_{d-1} \, r^{d-1} \, \partial_t \varphi \,
  \partial_r \varphi$ is computed at $r = R_s$. Similar law in $d=0$
  dimensions includes additional flux~$J_{0}$ 
  coming from $r=0$, see Sec.~\ref{sec:limit-d-to-zero}.} which is  
satisfied in $d\geq 1$ and $d=0$ with relative precision better  
than $10^{-9}$ and $5\cdot 10^{-5}$, respectively.

To end up, we remark that noticeable numerical errors in parameters
$\omega$ and $E$ of oscillons come from the fact that these objects
continuously radiate, lose energy and interfere with the emitted
waves. To decrease the fluctuations, we averaged these  parameters
over~${P> 1}$ consecutive oscillation periods and marked the
sensitivity to $P$ with the errorbars in
Figs.~\ref{fig:om_E} and~\ref{fig:sae-compare}b. Numerical errors in
all other figures are smaller 
than the circle size. 

%%%%%%%%%%%%%%%%%%%%%%%%%%%%%%%%%%%%%%%%%%%%%%%

\section{A pedagogical example}
\label{sec:Mech_Example}
Let us illustrate the classical EFT using two weakly
coupled harmonic oscillators $\varphi_1 (t)$ and $\varphi_2 
(t)$ with frequency $\Omega$ and Hamiltonian 
\begin{equation}
  \label{couple_osc_action}
  {\cal H} = \frac12 (   \pi_{1}^2 + \pi_2^2 )   +  \frac{\Omega^2}{2}
  (\varphi_1^2 + \varphi_2^2)   +  \frac12 \sum_{\alpha \beta}
  \lambda_{\alpha \beta} \; \varphi_\alpha \varphi_\beta\;.
\end{equation}
Here $\pi_\alpha = \partial_t \varphi_\alpha$ are the canonical
momenta and $\lambda_{\alpha \beta} \ll \Omega^2$  are  
small couplings. One can regard Eq.~(\ref{couple_osc_action}) as a
crude discretization of the  classical field with values
$\varphi_{1}$, $\varphi_2$ at two 
spatial points and the last term in Eq.~(\ref{couple_osc_action})
representing gradient energy.

Explicit diagonalization of the Hamiltonian gives two
eigenfrequencies, 
\begin{equation}
  \label{osc_omega}
  \omega_{\pm}^2 = \Omega^2 + \frac12 \mathrm{tr}\, \lambda \pm \frac12
  \sqrt{(\mathrm{tr}\,\lambda)^2 - 4\det \lambda}\;,
\end{equation}
where $\mathrm{tr}\, \lambda$ and $\det \lambda$ are the trace and
determinant of $\lambda_{\alpha \beta}$. 

Let us derive this nontrivial formula using the classical EFT at $\lambda
\ll \Omega^2$. In this case the oscillators $\varphi_1$ and
$\varphi_2$ are almost independent, and the last term in
Eq.~(\ref{couple_osc_action}) enables slow energy transfer 
between them. For a start, we introduce action-angle variables for
every oscillator at $\lambda=0$, 
\begin{equation}
  \label{eq:15}
  \varphi_\alpha = \sqrt{\frac{2I_\alpha}{\Omega}}\,  \cos \theta_\alpha\;, \qquad
  \pi_{\alpha} = -\sqrt{2I_\alpha\Omega} \, \sin \theta_\alpha, \qquad
  h_\alpha \equiv ( \pi_\alpha^2 + \Omega^2\varphi_\alpha^2)/2  =
  \Omega I_\alpha\;.
\end{equation}
Here $\alpha = 1,\, 2$ and $I_\alpha$ and $\theta_\alpha$ characterize
oscillation amplitudes and phases, respectively. During the motion of
decoupled oscillators $I_\alpha$ would be time-independent and~${\theta_\alpha
= \Omega t +  \mbox{const}}$.

Now, consider small nonzero $\lambda$. We perform canonical
transformation~(\ref{eq:15}) in the full
Hamiltonian~(\ref{couple_osc_action}) and average its last
(subdominant) term over the period of fast oscillations. It is  
convenient to introduce the quantities 
\begin{equation}
  \Theta = (\theta_1 + \theta_2)/2 \qquad
  \mbox{and} \qquad
  \vartheta = (\theta_1 - \theta_2)/2\;.
\end{equation}
In the decoupling limit $\lambda=0$, the value of $\Theta$ grows
linearly in time and $\vartheta$ remains constant~--- these are
the analogs of $\theta$ and $\partial_i \theta$ in field theory. Hence,
we can average the subdominant term of the Hamiltonian over
$\Theta$ instead of time, and move ``slow'' $\vartheta$ and
$I_\alpha$ out of the averages, cf.\ Eq.~(\ref{mu_coeff}). We obtain the
effective Hamiltonian  
\begin{align}
  \label{mech_eff_action}
  & {\cal H}_{\mathrm{eff}}  = \Omega (I_1 + I_2) + (\lambda_{11}I_1 +
    \lambda_{22} I_2) / (2\Omega) + \lambda_{12}\sqrt{I_1 I_2} \,
    \cos(2\vartheta)/\Omega
\end{align}
describing long-time evolution of $I_{\alpha}$ and
$\vartheta$. Notably, ${\cal
  H}_{\mathrm{eff}}$ is invariant under the global U(1) symmetry
$\theta_\alpha \to \theta_{\alpha} + \gamma$ or~$\Theta \to
\Theta + \gamma$.  This means  
that the respective Noether charge~${N = I_1 + I_2}$ is conserved. 

As a consequence of charge conservation, the effective theory has a
set of stationary solutions with $\partial_t I_{\alpha} = 
\partial_t \vartheta  = 0$ and $\Theta = \omega t$. Indeed,
one can check that this Ansatz passes the effective Hamiltonian
equations. Such solutions are the eigenmodes with
frequency~$\omega$ in the mechanical system. Also, they are the direct
analogs of oscillons in field theory.   

\begin{sloppy}
Instead of solving the effective equations, we can extremize the energy at a fixed $N$,
i.e., optimize the functional $F = {\cal H}_{\mathrm{eff}} - \omega N$
with Lagrange multiplier  $\omega$,
cf.\ Eq.~(\ref{F_generic}). Zero derivative with respect to $\vartheta$ is achieved at $\vartheta = 0$ or~$\pi/2$, while
extremization over $I_1$ and~$I_2$ gives,
\begin{equation} 
  \label{I_mech_eq}
  2\Omega (\omega -\Omega )\sqrt{I_1} = 
  \lambda_{11}\sqrt{I_1} \pm  \lambda_{12} \sqrt{I_2}\!\quad \mbox{and}\quad\!\!
  \pm 2\Omega (\omega-\Omega )\sqrt{I_2}  = \lambda_{12}
  \sqrt{I_1} \pm\lambda_{22}\sqrt{I_2}
  \,.
\end{equation}
Here the upper and lower signs correspond to $\vartheta = 0$ and
$\pi/2$, respectively. Equations~(\ref{I_mech_eq}) mean
  that $\psi_\alpha = \{\sqrt{I_1},\; \pm \sqrt{I_2}\}$ is an
  eigenvector of the matrix~$\lambda_{\alpha \beta}$ with  an
  eigenvalue~\mbox{$2\Omega (\omega-\Omega)$}. Solving this $2\times 2$
  eigenvalue problem, we obtain,
\begin{equation}
  \label{osc_omega_eff}
  \omega_{\pm} = \Omega + \frac{1}{4\Omega} \left[ \mathrm{tr}\,
    \lambda \pm \sqrt{(\mathrm{tr}\,\lambda)^2 - 4\det
      \lambda} \right]\;,
\end{equation}
where the $\pm$ signs discriminate between two eigenvalues; they are
  unrelated to those in Eq.~(\ref{I_mech_eq}). This result
coincides with Eq.~(\ref{osc_omega}) up to corrections of order 
$(\omega - \Omega)^2 = O(\lambda^2)$. We conclude that the the EFT works
at~${\lambda \ll \Omega^2}$. Once the stationary solutions
are found, the Hamiltonian equation $\partial_t \Theta =
\partial {\cal H}_{\mathrm{eff}} / \partial N = \omega$ tells us
that~$\Theta$ changes linearly in time, indeed. 
\end{sloppy}

Next, let us illustrate higher-order EFT corrections in the 
  mechanical system~(\ref{couple_osc_action}). To this end we
  perform canonical transformation~(\ref{eq:15}) in the full
  Hamiltonian. Besides, we break   $I_\alpha$ and $\theta_\alpha$ into 
  slow-varying and fast-oscillating  parts: 
\begin{equation}
	\label{mech:split_smooth_fast}
	I_\alpha = \bar{I}_\alpha + \delta I_\alpha\;, \qquad\qquad
	\theta_\alpha = \bar{\theta}_\alpha + \delta \theta_\alpha\;, 
\end{equation}
where $\langle \delta I_\alpha \rangle = \langle \delta \theta_\alpha
\rangle = 0$, ${\langle I_\alpha \rangle = \bar{I}_\alpha}$, and
${\langle \partial_t \theta_{\alpha} \rangle = \partial_t
  \bar{\theta}_{\alpha}}$. Equations for perturbations are obtained by
writing down the exact Hamiltonian equations
for~$I_\alpha$ and~$\theta_\alpha$ and subtracting their time
averages. We obtain ${\partial_t\delta 
I_{\alpha} = j_{\theta,\, \alpha}}$ and~${\partial_t \delta \theta_{\alpha} 
= - j_{I,\,   \alpha}}$, where 
\begin{equation}
	\label{mech:eq_corr}
	j_{\theta,\, \alpha}= - \sum\limits_{\beta} \lambda_{\alpha\beta} \left[
	\varphi_\beta \frac{\partial \varphi_\alpha}{\partial \theta_\alpha}
        - \Big\langle 
	\varphi_\beta \frac{\partial \varphi_\alpha}{\partial \theta_\alpha}\Big\rangle
	\right],\;\;
	j_{I,\, \alpha} =  -\sum\limits_{\beta} \lambda_{\alpha\beta} \left[
	\varphi_\beta  \frac{\partial \varphi_\alpha}{\partial
          I_\alpha} - \Big\langle 
	\varphi_\beta  \frac{\partial \varphi_\alpha}{\partial I_\alpha}\Big\rangle\right]
\end{equation}
are the sources, like in field theory,
cf.\ Eqs.~(\ref{eq_corrections}) and~(\ref{eq:41}). Recall that the
function~$\varphi_{\alpha}(I_{\alpha},\, \theta_\alpha)$ is given by
Eq.~(\ref{eq:15}) and note that~$j_{I,\,\alpha}$,~$j_{\theta,\,\alpha} \sim O(\lambda)$.

Now, we make the EFT approximations. First, we replace $I_{\alpha}$ and
  $\theta_{\alpha}$ in the sources with their leading-order parts
  $\bar{I}_{\alpha}$ and $\bar{\theta}_\alpha$. Second, we write the time
  derivative as~$\partial_t \approx (\partial_t \bar{\Theta}) \, 
  \partial_{\bar{\Theta}}$, since $\bar{\Theta} = (\bar{\theta}_1 + 
  \bar{\theta}_2)/2$ evolves almost linearly. This gives
  solutions for perturbations
  $\delta I_{\alpha} = {\cal I}[j_{\theta,\, \alpha}]/\partial_t \bar{\Theta}$ and~${\delta
  \theta_\alpha = -{\cal I}[j_{I,\, \alpha}]/\partial_t \bar{\Theta}}$,
  where
  \begin{equation}
    \label{eq:59}
          {\cal I}[f] = \int_0^{\bar{\Theta}} f(\bar{\Theta}') \,d\bar{\Theta}'  - \Big\langle
  \int_0^{\bar{\Theta}} f(\bar{\Theta}') \, d\bar{\Theta}'
  \;\Big\rangle 
  \end{equation}
  is a primitive on the $\bar{\Theta}$ circle and $\langle \cdot \rangle$ now
  averages over this variable, cf.~Eqs.~\eqref{delta_I_theta_solut}
  and~\eqref{I_primitive}. Explicitly, 
  \begin{equation}
    \label{eq:60}
    \delta I_\alpha = - \sum_\beta \frac{\lambda_{\alpha\beta}
      \sqrt{\bar{I}_{\alpha} \bar{I}_\beta}}{2\Omega\, \partial_t \bar{\Theta}} \;
    \cos(\bar{\theta}_\alpha + \bar{\theta}_{\beta})\;, \qquad
    \delta \theta_\alpha =  \sum_\beta \frac{\lambda_{\alpha\beta}
      \sqrt{\bar{I}_\beta}}{2\Omega\, \partial_t \bar{\Theta} \sqrt{\bar{I}_\alpha}} \;
    \sin(\bar{\theta}_\alpha + \bar{\theta}_{\beta})\;,
  \end{equation}
  where we substituted Eq.~(\ref{eq:15}) into
  Eqs.~\eqref{mech:eq_corr} and performed the
  $\bar{\Theta}$~integrals.

  The last step of our method is to expand the full Hamiltonian in
  perturbations to the second order in $\lambda$, substitute the
  solutions~\eqref{eq:60} and average the resulting expression over
  $\bar{\Theta}$. We obtain $\mathcal{H}_{\text{eff}} =
  \mathcal{H}^{(1)}_{\text{eff}} + \mathcal{H}^{(2)}_{\text{eff}}$,
  where $\mathcal{H}^{(1)}_{\text{eff}}$ is given by
  Eq.~\eqref{mech_eff_action} and  the second-order correction is
  \begin{equation}
    \label{eq:61}
    \mathcal{H}^{(2)}_{\mathrm{eff}} = - \frac{2}{\partial_t
      \bar{\Theta}} \sum_{\alpha} \langle\, j_{I,\, \alpha}\,  {\cal
      I}[j_{\theta,\, \alpha}]\,\rangle = - \sum_{\alpha \beta}
    \frac{(\lambda^2)_{\alpha \beta} \sqrt{\bar{I}_{\alpha}
        \bar{I}_{\beta}}}{4 \Omega^2 \, \partial_t \bar{\Theta}} \;
    \cos(\bar{\theta}_\alpha - \bar{\theta}_\beta)\;.
  \end{equation}
  In the last equality $\lambda^2$ is a square of the matrix
  $\lambda$.

  Note that in the main text we relied on the Lagrangian language, whereas this
  Appendix uses the Hamiltonian. This is
  equivalent: the second-order effective Lagrangian equals
  \begin{equation}
    \label{eq:62}
    {\cal L}_{\mathrm{eff}} = \sum_{\alpha} (\, \bar{I}_\alpha \partial_t
    \bar{\theta}_\alpha \,)- {\cal H}_{\mathrm{eff}}^{(1)} - \frac12
        {\cal H}_{\mathrm{eff}}^{(2)} + O(\lambda^3)\;,
  \end{equation}
  as one can see by performing the Legendre transform with respect to
  $\partial_t \bar{\theta}_\alpha$. 

\begin{sloppy}

  We test the second-order Hamiltonian by calculating the
  eigenfrequencies to the next order in $\lambda$. Note that ${\cal
    H}_{\mathrm{eff}}^{(1)} + {\cal H}_{\mathrm{eff}}^{(2)}$ is still
  invariant under the global symmetry $\bar{\Theta} \to \bar{\Theta} +
  \gamma$ because we averaged it over this variable. The respective
  global charge equals
  \begin{equation}
    \label{mech:N_corr}
    N  \equiv  \sum_\alpha\frac{\partial {\cal
        L}_{\mathrm{eff}}}{\partial (\partial_t 
      \bar{\theta}_{\alpha})}
    = \bar{I}_1 + \bar{I}_2 +
    \frac{\mathcal{H}^{(2)}_{\text{eff}}}{2\partial_t\bar{\Theta} }  +
    O(\lambda^3).
  \end{equation}
  As before, we find the eigenfrequencies by optimizing the
  functional ${F = \mathcal{H}_{\text{eff}} 
  - \omega N}$. Extremum with respect to $\partial_t \bar{\Theta}$ and
  $\vartheta$ is achieved at $\partial_t \bar{\Theta} = \omega$ and
  $\vartheta = 0$ or $\pi$. Varying over~$\bar{I}_1$ and $\bar{I}_2$, we
  obtain the eigenvalue problem,
  \begin{equation}
    2\Omega (\omega - \Omega) \psi_\alpha = \sum_{\beta}\left(\lambda
    - \frac{\lambda^2}{4\omega \Omega} \right)_{\alpha \beta}
    \psi_\beta + O(\lambda^3 \psi)\;,
  \end{equation}
  where again $\psi_{\alpha} \equiv \{ \sqrt{\bar{I}_1},\, \pm
  \sqrt{\bar{I}_2}\}$. We see that $\psi_\alpha$ is still an eigenvector of
  $\lambda$,  but the eigenfrequency $\omega_{\pm}$ receives a correction
  \begin{equation}
    \label{eq:63}
    \omega_{\pm}^{(2)} = - \frac{1}{32 \Omega^3} \left(\mathrm{tr}\,
      \lambda \pm \sqrt{(\mathrm{tr}\, \lambda)^2 - 4\det\lambda} \right)^2
  \end{equation}
  that should be added to
  Eq.~(\ref{I_mech_eq}). Equation~\eqref{eq:63} is precisely the
  $O(\lambda^2)$ term in the Taylor expansion  of the exact
  eigenfrequency \eqref{osc_omega}.

\end{sloppy}

It is worth stressing that the EFT
  series in~$\lambda$ converge, as is already clear from the exact
  result~(\ref{osc_omega}). This distinguishes our mechanical model  from the field theory case: nonperturbative effects
  like the ones governing the decay of the field-theoretical oscillons do not
  appear. It could have been expected, since our system of harmonic oscillators admits exactly periodic
  motions.

Note also that although we performed mechanical calculations in the
simplest illustrative model,  classical EFT may be helpful for
studying more involved dynamical systems. Indeed, this approach can be
easily applied to essentially nonlinear models, e.g.\ to several weakly
coupled anharmonic oscillators with slow dependencies of frequencies
on amplitudes. Traditional treatment of the 
latter system relies on KAM theory~\cite{Arnold} which is not easy to use.

%%%%%%%%%%%%%%%%%%%%%%%%%%%%%%%%%%%%%%%%%%%%%%%

\section{One-dimensional oscillons}
\label{sec:one-dimens-oscill}
In one dimension, the mechanical equation (\ref{eq:16}) for the oscillon
profile has a ``conservation law,'' i.e.\ $r$-independent
quantity 
\begin{equation}
  \label{E_1D}
  {\cal E} = \frac12 (\partial_r \chi)^2 + U_\omega(\chi) = \frac{2\psi^2}{\mu_I}\,
  (\partial_r \psi)^2 + \omega \psi^2 - h\;, \qquad \qquad
  \partial_r {\cal E} = 0\;,
\end{equation}
where $r = |x|$ and we transformed back to $\psi(r)$ using
Eqs.~\eqref{eq:26} and~\eqref{F_remaining}. Recall that the
form factors $\mu_I$ and $h$ in Eq.~(\ref{E_1D}) depend on $\psi^2$. We
are interested in the localized solutions approaching $\psi \to 0$ as
$r\to +\infty$. Hence,  ${\cal E} = 0$. Using Eq.~(\ref{E_1D}), we
immediately find,
\begin{equation}
  \label{eq:27}
  r = \int_\psi^{\psi_0}\frac{\psi' d\psi' \sqrt{2}}{\sqrt{\mu_I (h -
      \omega \psi'^2) }}\;,
\end{equation}
where the amplitude $\psi_0$ in the center satisfies equation $h(\psi_0^2)/\psi_0^2 =
\omega$. In the model~(\ref{U_def}), we substitute $\mu_I$ and $h$
from Eqs.~\eqref{eq:12}, \eqref{mu_coeff_th} and obtain
Eq.~(\ref{1d_analytic}).

The main idea in calculating the charge $N$  and  energy $E$ of
one-dimensional oscillons is to change the integration variable in
Eqs.~\eqref{eq:4} and~\eqref{F_generic} from $r$ to $\psi$ and then
express~$\partial_r \psi$ using Eq.~(\ref{eq:27}). This gives,
\begin{equation}
  \label{eq:28}
  N = 2 \int_0^\infty dr \, \psi^2(r) = 2^{3/2} \int_0^{\psi_0}
  \frac{\psi^3d\psi}{\sqrt{\mu_I (h - \omega \psi^2)}}
\end{equation}
and similarly for $E$. Using the form factors in the
model~(\ref{U_def}), we arrive\footnote{A change of variables 
  $\psi^2=(2-2\omega)(1-u^2)$ simplifies the integrals.}
 at Eqs.~(\ref{eq:30}), \eqref{eq:29} from the main text.

 %%%%%%%%%%%%%%%%%%%%%%%%%%%%%%%%%%%%%%%%%%%%%%%
\section{Vakhitov-Kolokolov criterion}
\label{sec:vakh-kolok-crit}
In this Appendix we prove the Vakhitov-Kolokolov criterion (\ref{eq:25})
which, if broken, indicates instability of oscillons under small
long-range perturbations. We will use the leading-order effective theory and energetic
argument of~\cite{vk,Zakharov12}. The latter applies without essential 
modifications, and this is nontrivial, since EFT is an unusual theory with 
non-canonical gradient term for~$\theta$ and nonlinear dependence of
the charge $N$ on the second, canonically normalized field~$\chi$, see
Eqs.~(\ref{action_eff_I}) and~(\ref{F_remaining}). 
Let us find out whether the oscillon solution $\chi =
\chi(\boldsymbol{x})$ and $\theta = \omega t$ truly minimizes the energy~$E$
at a fixed $N$, or it is just an unstable extremum. We  add small 
perturbations~$\delta \chi(\boldsymbol{x})$ and~$\delta
\theta(\boldsymbol{x})$ to the oscillon fields in such a way that $N$
in Eq.~(\ref{eq:4}) remains unchanged:
\begin{equation}
  \label{eq:33}
  \delta N = 0 \approx  \int d^d \boldsymbol{x} \, \frac{dI}{d\chi} \, \delta
  \chi = (\nu | \delta \chi ) \;,
\end{equation}
where we omitted the higher-order terms in $\delta \chi$, introduced 
$\nu(\boldsymbol{x}) = dI/d\chi = \sqrt{\mu_I}$ in accordance with
Eq.~\eqref{eq:26}, and defined the real scalar product ${(f | g ) = \int d^d
  \boldsymbol{x} \, f(\boldsymbol{x})\,
  g(\boldsymbol{x})}$. Hereafter all coefficient functions,
e.g.\ $\mu_I(\boldsymbol{x}) \equiv \mu_I(\chi(\boldsymbol{x}))$, are
evaluated on the background oscillon~$\chi(\boldsymbol{x})$. Note that
the norm of~$\nu(\boldsymbol{x})$  is finite at finite charge $N$,
since ${\nu^2 = \mu_{I} \propto  \psi^2(\boldsymbol{x})}$ in the
weak-field region $|\boldsymbol{x}|\to +\infty$, see 
Eqs.~(\ref{eq:4}) and~\eqref{eq:35}.

At a fixed $N$, variations of energy and $F\equiv E-\omega N$
coincide. We obtain,  
\begin{equation}
  \label{eq:34}
  \delta E = \delta F \approx \frac12 ( \delta \chi | \hat{L}_{\chi}| \delta
  \chi ) + \frac12 (\delta \theta | \hat{L}_{\theta} |\delta \theta)\;, 
\end{equation}
where Eqs.~\eqref{F_generic}, \eqref{eq:26}, and \eqref{F_remaining}
were used,
${\hat{L}_{\chi}  = -\Delta  -  U_\omega''(\chi(\boldsymbol{x}))}$
and~${\hat{L}_{\theta} = - \partial_i \, \mu_\theta^{-1}
  \,\partial_i}$ are the differential operators, and the primes denote
$\chi$ derivatives. Recall that $\mu_{\theta} \geq 0$, and hence
$\hat{L}_{\theta}$ is positive-definite,
cf.\ Eq.~\eqref{mu_coeff}. This means that the oscillons are  
stable if $\hat{L}_{\chi}$ is positive-definite in the
subspace~(\ref{eq:33}) of perturbations orthogonal to $\nu$.

It is remarkable that at least one eigenvalue of $\hat{L}_{\chi}$ is
negative, and the only question is whether it survives projection onto
the subspace~(\ref{eq:33}). Indeed, taking the $r$ derivative of
the profile equation~(\ref{eq:16}), 
one gets $(\partial_r \chi   
  | \hat{L}_\chi | \partial_r \chi ) = - (d-1) (\partial_r \chi | r^{-2}
  | \partial_r \chi) {< 0}$ in~${d>1}$ dimensions. The same
argument in $d=1$ implies that  $\partial_x \chi(x)$ is
a zero vector:~$\hat{L}_{\chi}|\partial_x  \chi) = 0$.  But this function also
has a node, $\partial_x \chi =0$ at $x=0$, thus implying by the oscillation
theorem that the eigenvector with smaller~--- negative~---
eigenvalue of $\hat{L}_\chi$ exists. We conclude that in any dimension
$\hat{L}_{\chi}$ has a negative eigenvalue. In what follows we will
use the orthonormal basis of $\hat{L}_{\chi}$ eigenvectors $\hat{L}_{\chi}
| n ) = w_n | n) $,  where $w_n$ are the eigenvalues and~${w_0 < 0}$. 

In the subspace (\ref{eq:33}) orthogonal to $|\nu)$ the eigenvalue problem for
$\hat{L}_{\chi}$ has a different form:
\begin{equation}
  \label{eq:36}
  \hat{L}_{\chi} | \delta \chi )  - C |\nu) = w |\delta \chi)\;.
\end{equation}
Here $w$ is the new eigenvalue and $C = (\nu | \hat{L}_{\chi}|
\delta \chi ) / (\nu| \nu)$ ensures that the left-hand side is
orthogonal to $|\nu)$. Equation~\eqref{eq:36} can be solved in the
full orthonormal basis of $\hat{L}_{\chi}$ eigenvectors:~$|\delta
\chi) = \sum_n \delta \chi_n | n)$ and $\delta \chi_n = C\, (n
| \nu ) / (w_n - w)$.

Once this is done, the eigenvalues $w$ in the subspace (\ref{eq:33})
are obtained by imposing condition~$(\nu | \delta \chi) = 0$. We
find,
\begin{equation}
  \label{eq:37}
  G(w) \equiv {\sum_n}' (n|\nu)^2 / (w_n - w)  = 0\;.
\end{equation}
Note that $\hat{L}_\chi$ has zero modes $\partial_i
\chi(\boldsymbol{x})$ representing shift symmetry: derivative of
Eq.~(\ref{eq:16}) gives~${\hat{L}_\chi | \partial_i \chi ) = 0}$. But 
they do not contribute into Eq.~(\ref{eq:37}) because $(\partial_i
\chi|\nu) = 0$ as the integral of the full derivative. Besides, in generic
case there are no other zero modes of~$\hat{L}_{\chi}$. Thus, in
Eq.~(\ref{eq:37}) and below we sum over all~$\hat{L}_\chi$ eigenmodes
except for the ones with zero eigenvalues and equip the respective sums
with primes. 

By construction, the function $G(w)$ grows, $\partial_w G > 0$, and
has poles at the original $\hat{L}_{\chi}$ spectrum $w=w_n$. In
particular, $G\to -\infty$ as~$w$ approaches $w_0 < 0$ 
from the above. We conclude that the solution of Eq.~\eqref{eq:37}
exists at $w<0$ if $G(0)>0$, i.e., 
\begin{equation}
  \label{eq:38}
  0 < G(0) = {\sum_n}'\, \frac{(\nu | n) (n | \nu)}{w_n} = (\nu |
  \hat{L}_\chi^{-1} | \nu)\;,
\end{equation}
where the ambiguity of $\hat{L}_{\chi}^{-1}$ in the linear span of zero
eigenvectors is removed by the matrix element with $|\nu)$. 
It is worth recalling that this condition guarantees negative
eigenvalue $w<0$ for
$\hat{L}_{\chi}$ in the subspace~(\ref{eq:33}) and hence indicates
instability of the background oscillon. 

To simplify Eq.~(\ref{eq:38}), we take a 
derivative of the profile equation~\eqref{eq:16} with respect to the 
oscillon frequency $\omega$. This gives~${\hat{L}_{\chi}\partial_\omega
\chi  = dI / d\chi = \nu(\boldsymbol{x})}$ or, inverting, $\hat{L}^{-1}_{\chi}|\nu )  =
|\partial_\omega \chi)$. Substituting this last equality into
Eq.~(\ref{eq:38}) and expressing the latter in terms of $dN / d\omega$, we get
the opposite of Eq.~(\ref{eq:25}), i.e.\ the condition for instability
of oscillons.

%%%%%%%%%%%%%%%%%%%%%%%%%%%%%%%%%%%%%%%%%%%%%%%

\section{Second-order effective action}
\label{Append:Corrections_F}

%%%%%%%%%%%%%%%%%%%%%%%%%%%%%%%%%%%%%%%%%%%%%%%

\subsection{Generalities}
\label{Append:Corrections_F:Gen}
In the EFT, we Taylor-expand Eq.~(\ref{aver_action}) with
respect to the perturbations $\delta I$ and~$\delta \theta$ and
average it over period. This strategy follows from the fact that the
time integral in the action kills mixing between slow and fast quantities,
e.g.,\ $\int dt \, \bar{I}\, \partial_t \delta \theta \approx \int dt
\,   \bar{I} \langle \partial_t \delta \theta \rangle = 0$. We will
extensively use this simplification below. The term~$\Phi \Delta \Phi$
is problematic, however, because $\Phi$ itself oscillates
with~$\bar{\theta}$. So, before expanding the  
Lagrangian we reorganize it: integrate by parts with respect to time,  express
$\partial_t\delta I$ and $\partial_t \delta \theta$  from 
the exact  equations for perturbations (\ref{eq_corrections}), and then add
and subtract the term~$\Phi \Delta \bar{\Phi}/2$, where~${\bar{\Phi} \equiv \Phi
(\bar{I}, \, \bar{\theta})}$. This gives, 
\begin{equation}
  \label{eq:F_tricked}
  {\cal S} = \int dt  d^d {\boldsymbol{x}} \left[  \bar{I}\partial_t \bar{\theta} +
    \frac12 \Phi \Delta \bar{\Phi} - h + \frac12 \Omega \delta I
    + \frac{\Delta \Phi}{2} \left(\Phi - \bar{\Phi} - \partial_I \Phi
    \, \delta I - \partial_{\theta} \Phi \,  \delta \theta\right)\right]\,,
\end{equation}
where all quantities without the overbar still depend on $I = \bar{I} + \delta
I$ and $\theta = \bar{\theta} + \delta \theta$. 

In the second-order EFT, we work to the quadratic order in $\delta I$
and $\delta \theta$. The bracket in the last term of
Eq.~(\ref{eq:F_tricked}) is already quadratic~--- hence, 
we can replace~$\Delta \Phi$ in front of it with $\Delta
\bar{\Phi}$. Expanding all terms, we obtain ${\cal S}  \approx {\cal
  S}_{\mathrm{eff}}^{(1)} + {\cal   S}^{(2)}_{\mathrm{eff}}$, where
\begin{equation*}
    \mathcal{S}_{\mathrm{eff}}^{(1)} = \int d t d^d\boldsymbol{x} \left[
      \bar{I}\partial_t \bar{\theta} - \bar{h}  + \frac{1}{2} 
      \bar{\Phi} \Delta \bar{\Phi}\right]\;, \qquad
    \mathcal{S}^{(2)}_{\mathrm{eff}} = 
    \frac{1}{2} \int dt d^d \boldsymbol{x}\left[ \Delta
      \bar{\Phi} \, \partial_{\bar{I}} \bar{\Phi}\, \delta I +
      \Delta \bar{\Phi} \, \partial_{\bar{\theta}} \bar{\Phi} \,
      \delta \theta \right]. 
\end{equation*}
The last step is to equip every term with the overall period average.
This makes~${\cal S}_{\mathrm{eff}}^{(1)}$ coincide with  the 
leading-order action~\eqref{mu_coeff},
(\ref{action_eff_I}). The correction ${\cal S}_{\mathrm{eff}}^{(2)}$
takes the form~\eqref{F_corr_gen} once the
solutions for perturbations~(\ref{delta_I_theta_solut})
and Eqs.~(\ref{eq:41}) are substituted\footnote{We also integrate by parts: $\langle {\cal I}[f]\,
  g\rangle = - \langle f\, {\cal 
    I}[g] \rangle$  if  $\langle f \rangle = \langle g \rangle =0$.}. 

Now, we write down the second-order effective action in the
case of oscillon: $\bar{\theta} = \omega t$ and~${\bar{I} =
\bar{I}(\boldsymbol{x})}$. It is convenient to introduce  
short-hand notations
\begin{equation}
  \label{eq:50}
  A(\bar{I}, \bar{\theta}\,) = \mathcal{I}\left[ \partial_{\bar{I}}
    \bar{\Phi} \, \partial_{\bar{\theta}} \bar{\Phi} \right]\;, \qquad \qquad 
  B(\bar{I}, \bar{\theta}) = (\partial_{\bar{I}} \bar{\Phi})^2 -
  \langle ( \partial_{\bar{I}} \bar{\Phi} )^2 \rangle\;.
\end{equation}
Some  combinations of higher $\bar{\Phi}$ derivatives can be
expressed via these quantities, e.g., ${\cal I} [\partial_{\bar{I}}^2
  \bar{\Phi} \, \partial_{\bar{\theta}} \bar{\Phi}] =
\partial_{\bar{I}} A - B/2$, where the property ${\cal
  I}[\partial_{\bar{\theta}} f] = f - \langle f \rangle$ 
was used. In terms of $A$ and $B$ the sources~(\ref{eq:41})
take the form,
\begin{equation}
  \label{eq:51}
  j_I =  B \,\Delta \bar{I} + \frac{1}{2} \partial_{\bar{I}} B\,
  (\partial_i \bar{I})^2\;, \qquad 
  \mathcal{I}[j_\theta] = A \, \Delta \bar{I} +
  \frac12 (2\partial_{\bar{I}} A - B)  (\partial_i \bar{I})^2\,.
\end{equation}
Substituting these expressions into the action~(\ref{F_corr_gen}), we
obtain the second-order correction~$-F^{(2)}$ to the Lagrangian,
\begin{equation}
  \label{eq:F_1_gen_form}
  F^{(2)} = - \int d^d {\boldsymbol{x}} \left[c_1 \,
    ( \partial_i \bar{I})^4 + c_2 \,
    (\partial_i \bar{I})^2 \Delta \bar{I} + c_3 \,
    (\Delta \bar{I})^2 \right]\;,
\end{equation}
where the form factors $c_i$ depend on $\bar{I}$:
\begin{align}
  \notag
  & c_1 = \frac{1}{4\omega}\, \big\langle \partial_{\bar{I}} B
  \,(2\partial_{\bar{I}} A - B)\big\rangle - \frac{\partial_{\bar{I}}
    \Omega}{8\omega^2} \, \big\langle (2\partial_{\bar{I}} A - B)^2
  \big\rangle\;,\\
  \label{eq:53}
  & c_2 = \frac{1}{2\omega} \, \big\langle
  A \partial_{\bar{I}} B + 2 B \partial_{\bar{I}} A - B^2
  \big\rangle - \frac{\partial_{\bar{I}} \Omega}{2\omega^2} \, \langle
  2 A\partial_{\bar{I}} A - AB \rangle\;,\\
  \notag
  & c_3 = \frac1{\omega} \, \big\langle AB \big\rangle -
  \frac{\partial_{\bar{I}} \Omega}{2\omega^2} \, \langle A^2 \rangle\;.
\end{align}
Notably, $c_i$ are quadratic in $A$ and $B$ and  hence involve four
$\bar{\Phi}$ multipliers in every term. We finally
change variables to $\psi(\boldsymbol{x}) = \sqrt{\bar{I}}$ and arrive to
Eq.~(\ref{eq:46}) with coefficients  
\begin{equation}
  \label{eq:48}
  d_1 = 16 \psi^4 c_1 + 8 \psi^2 c_2 + 4 c_3\;,  \qquad
  d_2 = 8(\psi^2 c_2 + c_3)\;, \qquad d_3 = 4 \psi^2 c_3\;,
\end{equation}
where the last formula coincides with Eq.~(\ref{eq:40}) once $c_3$ and
Eqs.~(\ref{eq:50}) are substituted.

%%%%%%%%%%%%%%%%%%%%%%%%%%%%%%%%%%%%%%%%%%%%%%%

\subsection{The model with a plateau potential}
\label{Append:Corrections_F:Explicit}
Next, we calculate the second-order EFT form factors $c_i$ and
$d_i$ in the particular model~(\ref{U_def}). It will be convenient to
use complex variable $z = - \exp(2i\bar{\theta})$ instead of $\bar{\theta}$;
now, the period averages are given by the integrals over unit circle
$|z|=1$, see Footnote~\ref{fn:1}. The function~$\Phi(\bar{I},\, z)$ in 
Eq.~\eqref{tanh_action_angle} has singularities at $z = z_1$ and 
$z=z_2$,
\begin{equation}
  \label{append:z_1_z_2_roots}
  z_1 = 2/\bar{I} - 1 \;, \qquad \qquad z_2 = \bar{I}/(2 - \bar{I}) = z^{-1}_1\;,
\end{equation}
which will  appear in all expressions. 

\begin{sloppy}

Substituting the expression~\eqref{tanh_action_angle} for $\Phi$
into Eqs.~\eqref{eq:50}, we find $A = N_A \xi_A$ and~${B = N_A 
  \xi_B}$, where we separated the $\bar{\theta}$  independent factor 
\begin{equation}
  \label{eq:56}
  N_A =\left[2\bar{I} (1-\bar{I}) (2-\bar{I}) \right]^{-1}
\end{equation}
and denoted
\begin{equation}
 \xi_A = \ln ( 1 - z_2/z) + \ln ( 1 -
   z/z_1)\;, \qquad \xi_B =
 -\frac{2}{\bar{I}(2-\bar{I})} \left( \frac{z}{z_1-z} + \frac{z_2}{z
     - z_2}\right)\;.
\end{equation}
It is remarkable that $\xi_{B} = \partial_{\bar{I}} \xi_A$. Using this
property, we can express all period averages in the
form factors~(\ref{eq:53}) in terms of quantities $\langle \xi_A^2
\rangle$ and $\langle (\partial_{\bar{I}} \xi_A)^2\rangle$ and their
$\bar{I}$ derivatives. Say, the third coefficient equals
\begin{equation}
  \label{eq:58}
  c_3 = \frac{N_A^2}{2\omega^2} \left( \omega \, \partial_{\bar{I}}
  \langle \xi_A^2 \rangle +  \langle \xi_A^2 \rangle
  \right)\;,
\end{equation}
and others have similar form; recall that $\Omega = 1-\bar{I}$ in
our model. Evaluating the contour integrals over $z$, we obtain, 
\begin{equation}
  \label{eq:57}
  \left\langle \xi^2_A \right\rangle = 2 \operatorname{Li}_2 (z^2_2)
  \qquad \text{and} \qquad \left\langle (\partial_{\bar{I}}\xi_A)^2
  \right\rangle = \frac{2}{(1-\bar{I})(2-\bar{I})^2}\;,
\end{equation}
where ${\mathrm{Li}_2(w) = -\int_0^w
  dw'\, \ln(1-w')/w'}$ is the dilogarithm. 

\end{sloppy}

The final result is,
\begin{align}
  \notag
  c_1 =  \; &\frac{\ln (1  - z_2^2)\, \partial_{\bar{I}} N_A^2}{\omega \bar{I}(2-\bar{I})}
                \left\{ \frac{2-2\bar{I}}{\bar{I}(2-\bar{I})}   - \frac{1}{\omega}
                  -  \frac{\partial_{\bar{I}} N_A}{N_A}\right\}
      +      \frac{(\partial_{\bar{I}} N_A)^2}{\omega^2} \;
                \mathrm{Li}_2(z_2^2) \\[2px]  \label{c_1_fin}
                \; &\qquad \qquad \qquad \qquad
                +     \frac{N_A^2}{4\omega (1 - \bar{I})(2 - \bar{I})^2} \, \left\{
               \frac{4-3\bar{I}}{(1 - \bar{I})(2-\bar{I})} + 6\,
               \frac{\partial_{\bar{I}} N_A }{N_A} +
               \frac1\omega\right\} \;,\\[4px] \notag
  c_2 =\; &  \frac{2\ln (1-z_2^2) \, N_A^2}{\omega \bar{I} (2-\bar{I})}
               \left\{\frac{2-2\bar{I}}{\bar{I}(2-\bar{I})}  - 
               \frac{3\partial_{\bar{I}} N_A}{N_A} - \frac{1}{\omega}\right\}
            + \frac{2 N^2_A}{\omega (1-\bar{I})(2-\bar{I})^2} 
            + \frac{\partial_{\bar{I}} N^2_A}{\omega^2}\;
   \mathrm{Li}_2(z^2_2)\;, \notag \\[4px]\notag
c_3 = \; &\frac{N^2_A}{\omega} \left\{
       \frac{1}{\omega} \, \mathrm{Li}_2\left(z^2_2\right)
       - \frac{4 \ln(1-z_2^2)}{\bar{I}(2-\bar{I})} \right\}\;,
       \notag
\end{align}
 where $z_2$ and $N_A$ are given by Eqs.~(\ref{append:z_1_z_2_roots}),
 (\ref{eq:56}). Now, $d_i$ can be found from Eqs.~(\ref{eq:48}).
 
%%%%%%%%%%%%%%%%%%%%%%%%%%%%%%%%%%%%%%%%%%%%%%

\bibliography{oscillon}{}
\bibliographystyle{JHEP}

\end{document}